 \documentclass[twocolumn]{aastex63}
\usepackage{graphics,epsfig,longtable,txfonts,color}
\usepackage[english]{babel}

\newcommand{\myemail}{hstiele@mx.nthu.edu.tw}

\newcommand{\mr}{\mathrm}

\def\eg{e.\,g.}                                      

\def\xmm{\textit{XMM-Newton}}
\def\swift{\textit{Swift}}

\def\gx339{GX\,339-4}
\def\h1743{H\,1743-322}

\def\max15{MAXI\,J1535--571}
\def\max18{MAXI\,J1820+070}

\def\aap{A\&A}

\def\apjl{ApJ}

\def\apjs{ApJS}

\shorttitle{2018/19 outbursts of \max18}
\shortauthors{Stiele, Kong}

\begin{document}


\title{A timing study of \max18\ based on \swift/XRT and NICER monitoring in 2018/19}

\author{H.\ Stiele}
\affil{Institute of Astronomy, National Tsing Hua University, No.~101 Sect.~2 Kuang-Fu Road,  30013, Hsinchu, Taiwan}
\email{\myemail}
\and

\author{A.\ K.\ H.\ Kong}
\affil{Institute of Astronomy, National Tsing Hua University, No.~101 Sect.~2 Kuang-Fu Road,  30013, Hsinchu, Taiwan}





\begin{abstract}
We present a detailed timing analysis of the bright black hole X-ray binary \max18\ (ASASSN-18ey), during its first detected outburst lasting from March 2018 until October 2019 based on \swift/XRT window timing mode observations, corresponding UVOT data and NICER observations. The light curves clearly show four outbursts, with the source remaining in the hard state during its first outburst, while the rise of the second outburst corresponds with the transition to the soft state. A similar double outburst of GX339-4 has been observed in 2004. Here it is followed by two hard-state only outbursts. In many observations the power density spectra showed type-C quasi-periodic oscillations with characteristic frequencies below 1~Hz, which suggests that the source stayed in a state of low effective accretion for large parts of its outburst. The absence of other types of quasi-periodic oscillations hinders a precise determination of the state transitions, but from combining NICER and \swift/XRT data, we find that \max18\ went from the hard-intermediate to the soft state in less than one day. The covariance ratios derived from NICER data show an increase towards lower energies, which indicate that the source should make a transition to the soft state. This transition finally took place, after \max18\ stayed in the hard state at rather constant luminosity for about 116 days. The steepness of the increase of the covariance ratios is not correlated with the amount of rms variability and it does not show a monotonic evolution along the outburst.
\end{abstract}


\section{Introduction}
Almost all low-mass black hole X-ray binaries (BHBs) are transient sources, which means that they show outbursts, which typically last from weeks to months. During their outbursts they evolve through different states \citep{2006csxs.book..157M,2010LNP...794...53B}. Using hardness intensity diagram \citep[HID;][]{2001ApJS..132..377H,2005A&A...440..207B,2005Ap&SS.300..107H,2006MNRAS.370..837G,2006csxs.book..157M,2009MNRAS.396.1370F,2010LNP...794...53B,2011BASI...39..409B}, hardness root-mean square (rms) diagram \citep[HRD;][]{2005A&A...440..207B} and rms intensity diagram \citep[RID;][]{2011MNRAS.410..679M} the evolution during the outbursts can be studied. At the beginning and end of an outburst the BHBs are found in the low-hard state (LHS). In this state an rms of several tens of per cent is observed and the emission is dominated by thermal Comptonization in a hot, geometrically thick, optically thin plasma located in the vicinity of the black hole, where softer seed photons coming from an accretion disk are up-Comptonized \citep[see][for reviews]{2007A&ARv..15....1D,2010LNP...794...17G}. Many BHBs show transitions to the high-soft state (HSS), in which the variability is much lower \citep[fractional rms $\sim$1 per cent, e.g.][]{2005A&A...440..207B} and in which the spectrum is clearly dominated by an optically thick, geometrically thin accretion disk \citep{1973A&A....24..337S}.

It is thought that the transitions reflect major changes in the properties of the inner accretion flow \citep[\eg\ ][]{2007A&ARv..15....1D}. A geometrically thin, optically thick accretion disk with an inner truncation radius varying as a function of the accretion rate can explain the different states and transitions between them \citep[\eg\ ][]{1997ApJ...489..865E,2001ApJ...555..483E}. While there is observational evidence that during the HSS the accretion disk extends down to the innermost stable circular orbit \citep[ISCO;][]{2004MNRAS.347..885G,2010ApJ...718L.117S}, the extension of the accretion disk in the LHS is still a matter of ongoing debate. Different observational studies give different estimates of how close the disk comes to the black hole in this state \citep[\eg\ ][]{2006ApJ...653..525M,2014A&A...564A..37P,2010MNRAS.407.2287D,2014MNRAS.437..316K,2015A&A...573A.120P}. The truncated disk model assumes that the disk recedes in the LHS and that the inner parts are filled by a radiatively inefficient, optically thin, advection-dominated accretion flow \citep[\eg\ ][]{1995ApJ...452..710N,1997ApJ...489..865E}.

Not all Galactic BHBs make the transition to the soft state and $\sim40$\% of them remain in the hard state during the entire outburst \citep{2016ApJS..222...15T}. This type of outburst has been dubbed ``failed'' outburst \citep{2009MNRAS.398.1194C,2016MNRAS.460.1946S}. The physical reason why some outbursts remain in the hard state, while other make it to the soft state, is a topic of ongoing research. Failed outbursts tend to have lower peak luminosities ($\la0.11L_{\mr{edd}}$) and they seem to occur more commonly in shorter orbital period systems \citep{2016ApJS..222...15T,2013MNRAS.434.2696S}. They might be connected to a premature decrease of the mass accretion rate \citep{2009MNRAS.398.1194C} . 

The different states also show different features in power density spectra \citep[PDS;][]{2014SSRv..183...43B}. In  a large number of sources type-C quasi-periodic oscillations (QPOs) \citep[][and references therein]{1999ApJ...514..939W,2011MNRAS.418.2292M} have been observed in the LHS and hard intermediate state \citep[HIMS;][]{2006csxs.book..157M,2010LNP...794...53B}. These oscillations have centroid frequencies ranging from 0.01 to 30 Hz, and their quality factor ($Q=\nu_0/(2\Delta)$, where $\nu_0$ is the centroid frequency, and $\Delta$ is the half width at half maximum) is $\ga10$ \citep[see \eg\ ][]{2005ApJ...629..403C,2010ApJ...714.1065R}. Often they appear with one or two overtones and at times with a sub-harmonic. The PDS always show  band limited noise and the characteristic frequency of the QPO is anti-correlated with the total broad-band fractional rms variability. These oscillations are thought to be caused by Lense-Thirring precession of a radially extended region of the hot inner flow \citep{1998ApJ...492L..59S,2009MNRAS.397L.101I}. The presence of type-B QPOs together with a weaker power-law noise defines the soft intermediate state (SIMS). Type-B QPOs  have centroid frequencies of 0.8 -- 6.4 Hz, $Q>6$, have a 5 -- 10\% fractional rms and appear often together with an overtone and a sub-harmonic. The third type of QPOs are type-A QPOs. They have centroid frequencies of 6.5 -- 8 Hz, are broad ($Q\sim1-3$) and weak (fractional rms $<5$\%).  The three QPO types are well separated as a function of the total integrated fractional rms in the power density spectrum.

In this paper, we present a comprehensive study of the temporal variability properties of the X-ray transient \max18\ observed during four outbursts between March 2018 and October 2019. On March 11, 2018 MAXI/GSC \citep{2018ATel11399....1K} detected a bright uncatalogued  hard X-ray transient. About half a day later the source was detected by \swift/BAT \citep{2018ATel11403....1K}. Based on follow-up observations in the radio, optical and X-ray band the source has been classified as a black hole X-ray binary candidate \citep{2018ATel11418....1B,2018ATel11420....1B}.  State transitions and the detection of quasi-periodic oscillations (QPOs) have been reported \citep{2018ATel11576....1H,2018ATel11578....1B,2018ATel11820....1H}. \citet{2019Natur.565..198K} found thermal reverberation lags that are shorter than those previously observed from black hole X-ray binaries. While the timescale of the reverberation lags shows some evolution, the shape of the broadened iron K emission line remains rather constant. These findings suggest that there is not much evolution in the truncation radius of the inner disk during the luminous hard state. A detailed analysis of the spectral properties of \max18\ during this outburst based on MAXI/GSC and \swift/BAT data has been reported \citep{2019ApJ...874..183S}. The results of a spectral study based on \swift/XRT and NuSTAR data taken a few days after the detection of the outburst have been presented \citep{2019MNRAS.487.5946B}. X-ray spectra and a light curve based on \xmm\ data obtained during outburst rise are presented in \citet{2019arXiv190606519K}. The light curve shows pronounced dipping intervals. In the same study the UV/X-ray cross correlation function was investigated. A detailed optical spectroscopic follow-up study of the optical counterpart ASASSN-18ey detected clear accretion disk wind features in the hard state \citep{2019ApJ...879L...4M}. Here, we investigate the timing properties of \max18\ using Neil Gehrels \swift\ Observatory/XRT and UVOT and NICER (Neutron star Interior Composition Explorer) data. More details on the data used and on the data analysis are given in Sect.~\ref{Sec:obs}. We present our results in Sect.~\ref{Sec:res} and discuss them in Sect.~\ref{Sec:dis}.

\section[]{Observation and data analysis}
\label{Sec:obs}
\subsection{Neil Gehrels \swift\ Observatory}
\label{Sec:obs_sw}
\subsubsection{XRT}
All \swift/XRT \citep{2005SSRv..120..165B} monitoring data of \max18\ obtained in window timing mode between 2018 March 11th and 2019 September 28th are analysed in this study. 
To obtain power density spectra (PDS) in the 0.3 -- 10 keV energy band, we make use of the GHATS package (v.~1.1.1), developed under the IDL environment at INAF-OAB \footnote{\url{http://www.brera.inaf.it/utenti/belloni/GHATS\_Package/Home.html}}. We follow the procedure outlined in \citet{2006MNRAS.367.1113B}, subtracting the contribution due to Poissonian noise \citep{1995ApJ...449..930Z}, normalising the PDS according to \citet{1983ApJ...272..256L} and converting to square fractional rms \citep{1990A&A...227L..33B}. We did not include any corrections of the PDS due to background photons. We determine the contribution due to Poissonian noise by fitting the flat tail of the PDS at the high-frequency end with a constant. This allows us to take into account deviation from the expected value of 2, that are caused by pile-up effects in the \swift/XRT data \citep{2013ApJ...766...89K}. To fit the PDS within \textsc{isis} \citep[v.~1.6.2;][]{2000ASPC..216..591H} we use zero-centered Lorentzians for band-limited noise (BLN) components, and Lorentzians for QPOs. 

\subsubsection{UVOT}
For all \swift/XRT monitoring observations we also analysed the UVOT data \citep{2005SSRv..120...95R}. This data have been pre-processed at the Swift Data Centre \citep{2010MNRAS.406.1687B} with the current HEAsoft version and current calibration files and we use the HEAsoft task \texttt{uvotproduct} to obtain magnitudes from the images in the different bands available for each observation. As source position we used R.A. = 18$^{\rm h}$20$^{\rm m}$21\fs9, dec. = +07$^{\circ}$11$^{\prime}$07\farcs3 \citep{2018ApJ...867L...9T} with an extraction radius of five arc-seconds. A nearby source free region was used to define a background region with a radius of ten arc-seconds. 

\subsection{NICER}
NICER \citep{2012SPIE.8443E..13G} observed \max18\ between 2018 March 12th and 2018 July 5th, where observations with an exposure of less than 1 ks have been excluded from our study. NICER resumed observing this source on 2018 September 25th and followed it until 2019 October 25th. (All NICER observations taken between 2018 July 6th and 2018 September 24th ares given with an exposure of zero seconds in the HEADAS data base.) We use the pre-processed event files provided by the NICER data center. These files are produced using the current HEAsoft version and current calibration files. We derive Poissonian noise subtracted, Leahy normalised and to square fractional rms converted PDS using HEAsoft (v.~6.23) tasks.  

\begin{figure*}
\resizebox{\hsize}{!}{\includegraphics[clip,angle=0]{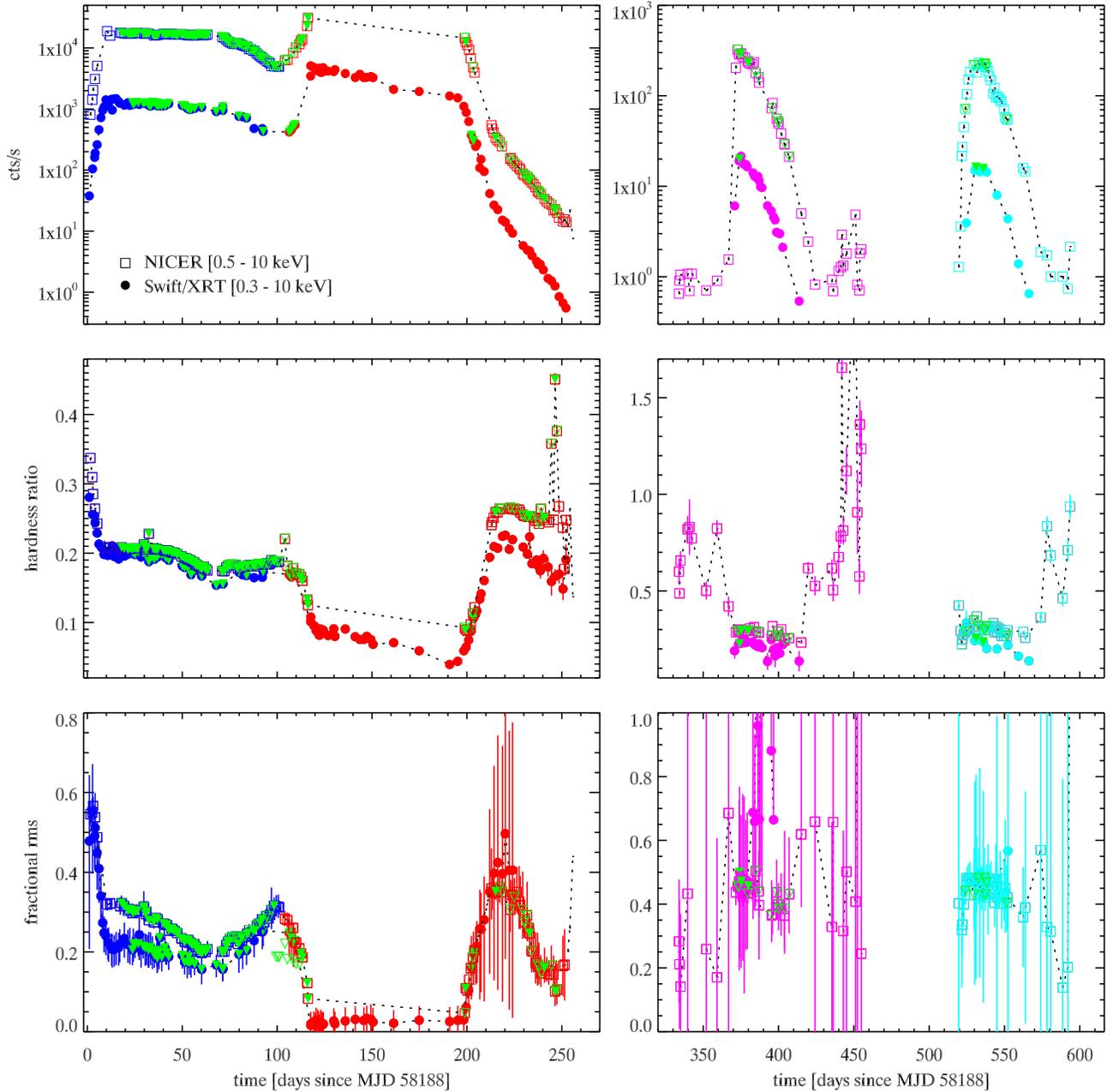}}
\caption{Light curves and time evolution of the hardness ratio and the fractional rms (from top to bottom) of the four outbursts of \max18\ between March 2018 and October 2019 based on Swift/XRT and NICER data. Each data point represents one observation. Observations of the four outbursts are indicated by different colours. The first two outbursts are displayed in the left hand panels, while the third and fourth outbursts are shown in the right hand panels. Please notice the change on the scales of the y-axis for corresponding panels. Observations in which a type-C QPO has been detected are marked by (green) triangles. In case of NICER observations filled  triangles indicate QPOs detected at $\ge3\sigma$, while open triangles indicate QPOs with a detection significance below $3\sigma$. T=0 corresponds to March 11th 2018 00:00:00.000 UTC.}
\label{Fig:LC}
\end{figure*}

\begin{figure*}
\resizebox{0.45\hsize}{!}{\includegraphics[clip,angle=0]{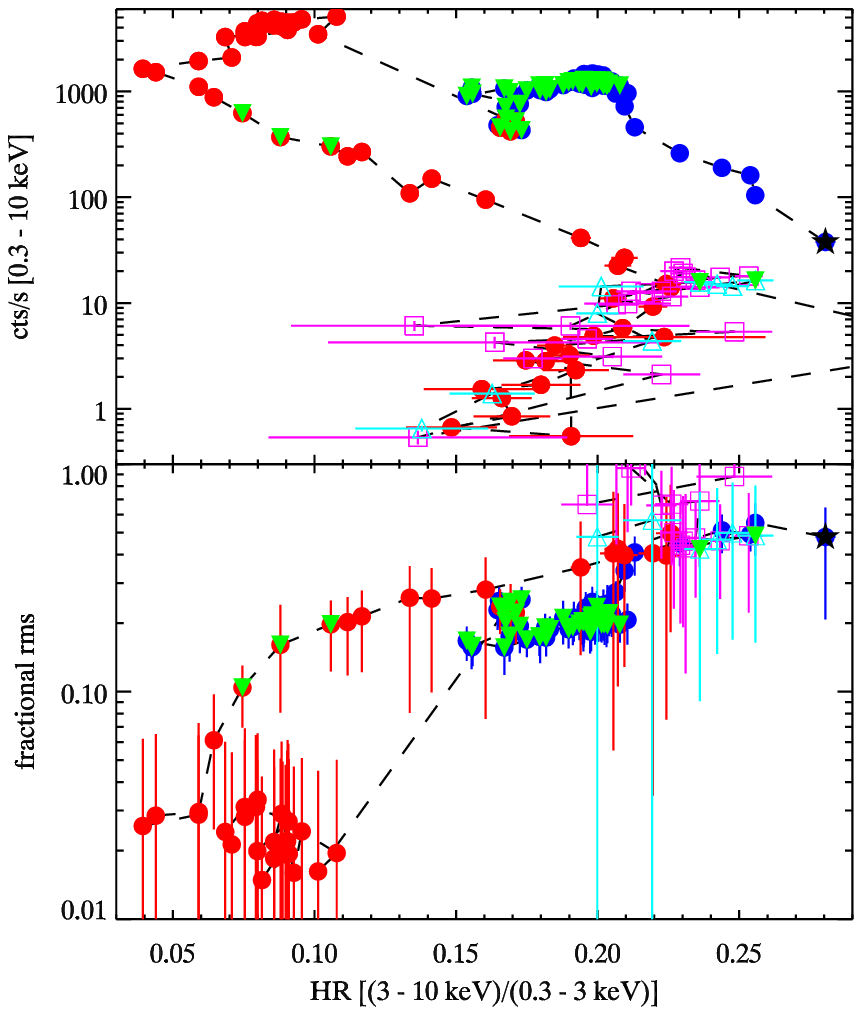}}\vspace{0.5cm}\resizebox{0.45\hsize}{!}{\includegraphics[clip,angle=0]{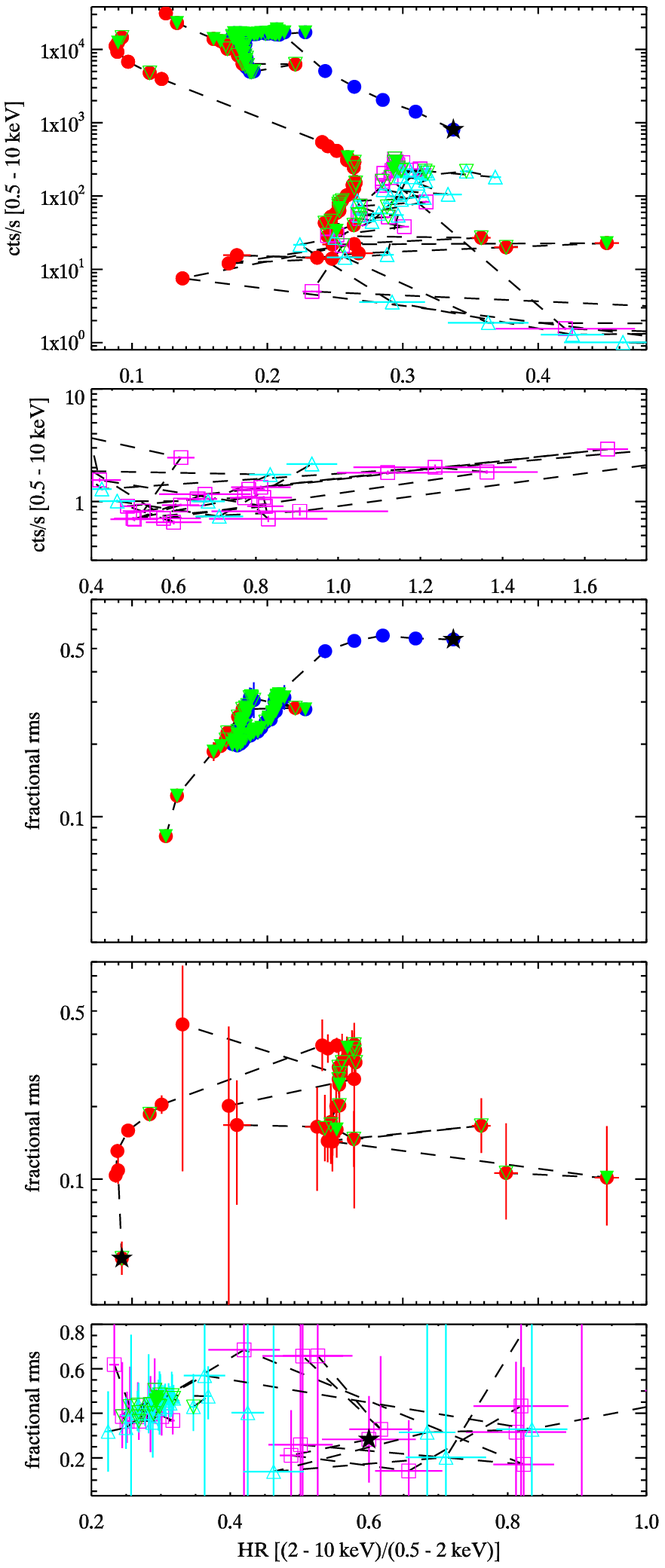}}
\caption{Hardness-intensity diagram (HID; upper panel) and hardness-rms diagram (HRD; lower panel), derived using \swift/XRT data (left panels). In case of NICER data (right panels) the upper two panels are the HID, where the lower one shows a higher HR range. These observations are not shown in the HRD (lower three panels; before the soft state, after the soft state, third and fourth outburst from top to bottom) as their rms values are not constrained, due to their low count rates. Filled and open down-pointing triangles indicate observations with a QPO detected above or below 3$\sigma$, respectively. Open squares indicate observations of the third outburst (after day 330), open up-pointing triangles observations of the fourth outburst (after day 500). The star marks the first observation. Observations of the three outbursts are indicated by different colours, following the colour scheme of Fig.~\ref{Fig:LC}. Each data point represents one observation. }
\label{Fig:HID}
\end{figure*}

\begin{figure}
\resizebox{\hsize}{!}{\includegraphics[clip,angle=0]{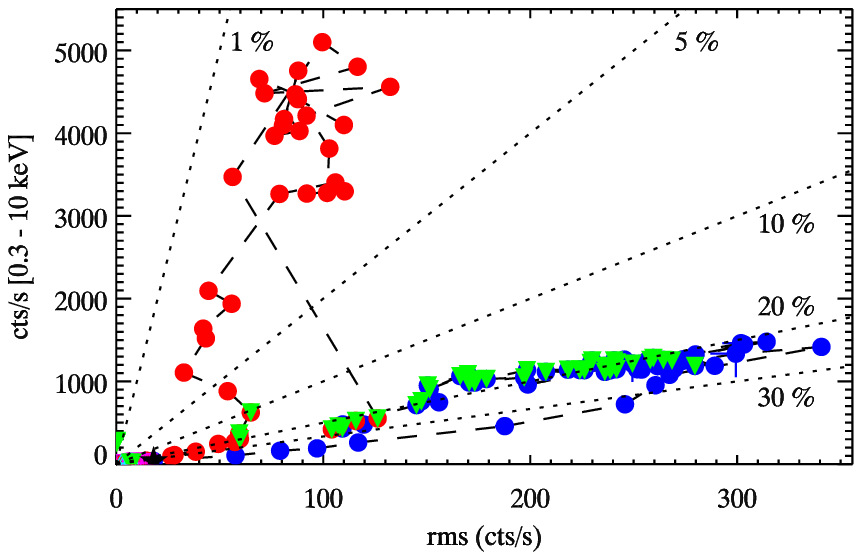}}\\ 
\smallskip
\resizebox{\hsize}{!}{\includegraphics[clip,angle=0]{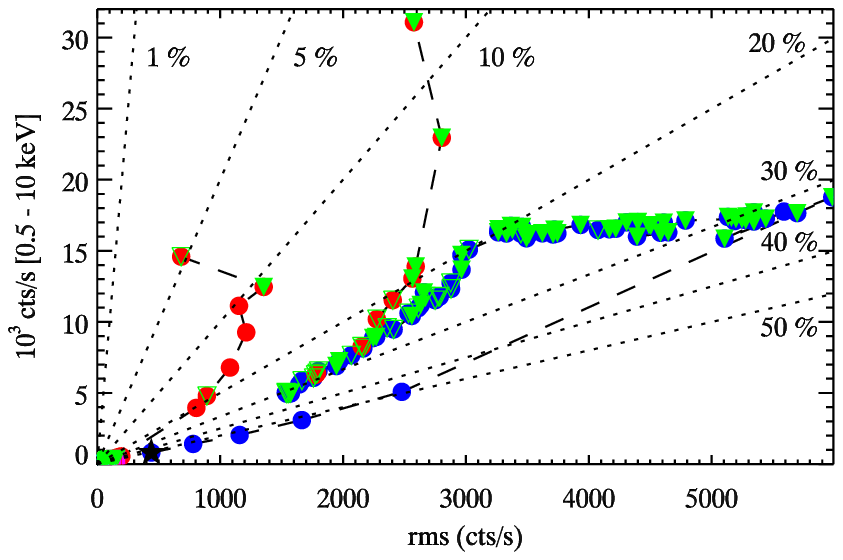}}\\ 
\smallskip
\resizebox{\hsize}{!}{\includegraphics[clip,angle=0]{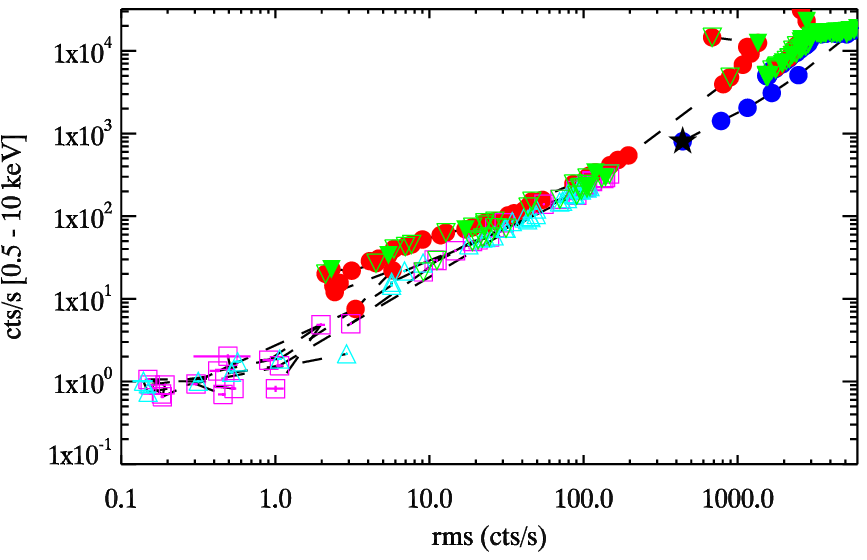}}
\caption{rms-intensity diagram, derived using \swift/XRT (upper panel) and NICER (lower two panels) data. While the middle panel uses linear axes, logarithmic axes are used in the bottom panel. Each data point represents one observation. Symbols and colours are the same as in Fig.~\ref{Fig:HID}.}
\label{Fig:RID}\
\end{figure}

\section[]{Results}
\label{Sec:res}

\subsection{Diagnostic diagrams}
\label{Sec:diag}
Based on \swift/XRT data we determined source count rates in three energy bands using the online data analysis tools provided by the Leicester \swift\ data centre\footnote{\url{http://www.swift.ac.uk/user\_objects/}} (which used HEAsoft v.6.22), including single pixel events only: total (0.3 -- 10 keV), soft (0.3 -- 3 keV), and hard (3 -- 10 keV). We also derived NICER count rates in the 0.5 -- 10 keV, 0.5 -- 2 keV and 2 -- 10 keV bands. The \swift/XRT and NICER light curves and time evolution of the hardness ratio (HR) and the fractional rms are shown in Fig.~\ref{Fig:LC}. To obtain HRs we divide the count rate observed in the hard band by the one obtained in the soft band. For \swift/XRT the fractional rms is determined in the 0.3 -- 10 keV band and in the $4\times10^{-3}$ -- 35.13 Hz frequency range, while for NICER we use the total energy band (0.0 -- 15.01 keV) and the $6.1\times10^{-3}$ -- 50 Hz frequency range. The hardness-intensity diagram (HID) and hardness-rms diagram (HRD) are shown in Fig.~\ref{Fig:HID}, while Fig.~\ref{Fig:RID} is the rms-intensity diagram (RID). 

The light curves show a steep rise lasting for about 15 days, before a plateau at a rather constant count rate of about $10^3$ ($10^4$) cts/s for \swift/XRT (NICER) is reached. This plateau lasts for about 55 days, before the count rate starts to decrease slowly, as can be seen from the NICER light curve, which provides a better coverage during this part of the outburst. After another 30 days the evolution of the light curves reverses and the count rate starts rising again for about 15 days before a second, slowly decaying plateau at a higher count rate of $3-4\times10^3$ cts/s for \swift/XRT is reached. This second plateau lasts for about 73 days. When the count rate drops to values comparable to the highest values observed during outburst rise (around day 10) the decay of the light curve steepens significantly. The NICER and \swift/XRT motoring end after 255.8 and 252.2 days, respectively. 78.2 days after the last NICER observation NICER resumed observing \max18. About 37 days after the beginning of these observations \max18\ shows another outburst that stays below the count rate at which the source was detected originally. The rise is again rather steep and it is directly followed by a decay similar to the one seen at the end of the first plateau. The long-term MAXI/GSC light curve shows that the count rate between the second and third outburst is similar to the pre-detection level. About 64.62 days after the last observation of the third outburst NICER resumed observing \max18\  and covered another outburst that reaches a similar count rate and has a similar shape as the third outburst.

The HID shows that the source softens during outburst rise and that the softening continues during the first plateau seen in the light curve until an HR of 0.15 (0.17) for \swift/XRT (NICER) is reached. The decrease in count rate takes place at a slightly higher HR of 0.17 (0.18--0.19). The second plateau has softer HRs, where most observations have an HR$<0.1$. The softest HR is obtained at the end of the second plateau. During the steep decay \max18\ hardens. Below NICER count rates of 200 cts/s the HR decreases slightly from 0.26 to 0.24 with decreasing count rate, before it hardens significantly at NICER count rates $< 30$ cts/s and softens again at $< 20$ cts/s. In observations taken before the third outburst at a NICER count rate $< 5$ cts/s the source has HR between 0.4 and 0.9.  The HRs observed during the third outburst are a little bit harder than the HRs observed towards the end of the second outburst, but softer than the HR at the beginning of the first outburst. Anyway, the source remained in the hard state during its third outburst. The HRs observed during the fourth outburst lie in a similar range as those observed during the third outburst, so this outburst remained again in the hard state.

From the HRD we see that the softening during outburst rise and first plateau correlates with a decrease in the fractional rms to about 15\% (in \swift/XRT data), and that the decrease in count rate takes place at higher rms values around 20\%. In the second plateau the fractional rms dropped significantly to values of about 2--3\%, although error bars are quite large at those low rms values. During the steep drop of the count rate, when the source hardens, we also observe an increase in rms, reaching values of 35\% in the NICER data. The  decrease in count rate below a NICER count rate of 200 cts/s corresponds to a decrease in rms from 35 to 14\%. The softening at $< 20$ cts/s corresponds to an increase in the rms from 10 to 20\%.

The RID shows the hard line at outburst rise, at fractional rms values similar to those seen in \gx339\ \citep{2011MNRAS.410..679M}. During the first plateau the fractional rms drops to about 20\% and then again increases slowly to 30\% while the total rms decreases, leading to a flat loop-like structure in the RID. During the second rise the total rms increases with increasing count rate, while the fractional rms decreases. The points of the second plateau lie in the region of the soft branch \citep{2011MNRAS.410..679M}. During the steep decay seen in the light curve, the increase in fractional rms corresponds to an increase of total rms, until the fractional rms gets close to 20\%, when the total rms starts to decrease and \max18\ follows again the hard line.

\subsection{UVOT light curves}
\begin{figure}
\centering
\resizebox{\hsize}{!}{\includegraphics[clip,angle=0]{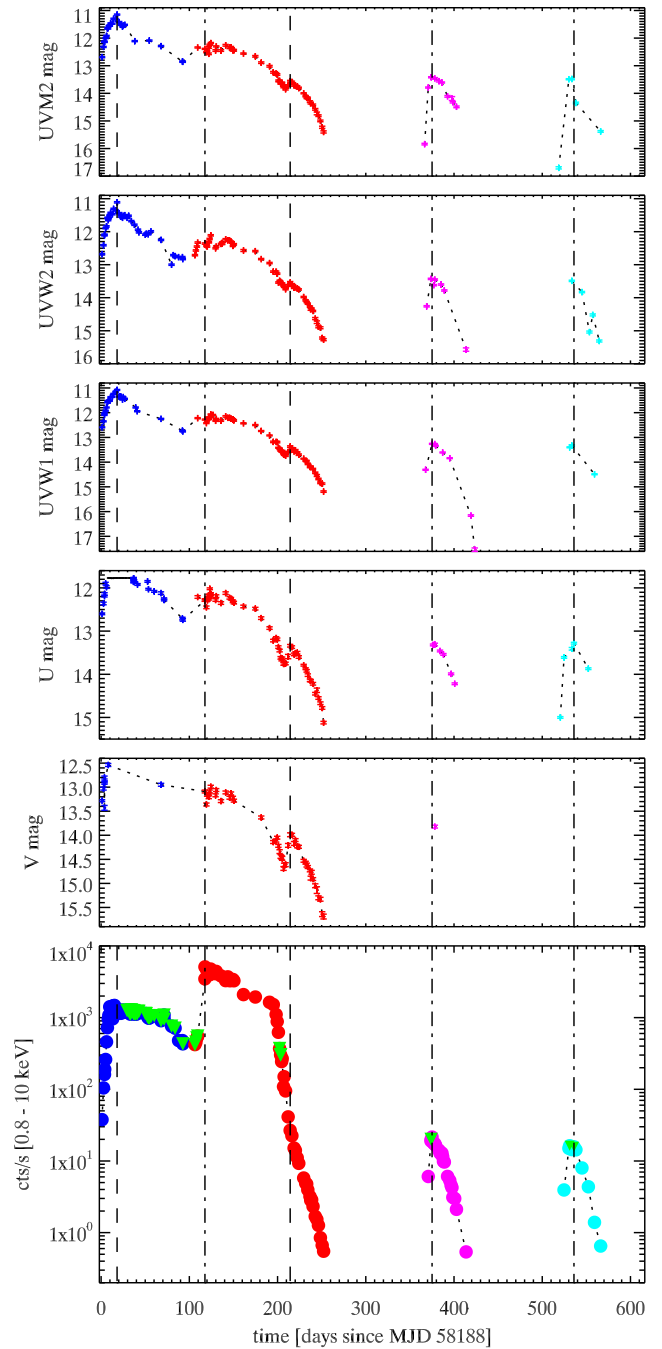}}
\caption{Light curve in the different UVOT bands. For comparison the \swift/XRT light curve is shown in the lowest panel. Each data point represents one observation. Observations of the four outbursts are indicated by different colours, following the colour scheme of Fig.~\ref{Fig:LC}. The U band observations in which the detector saturates when the source is at its brightest stage are not shown. The dashed lines indicate the time of the brightest magnitude in the UV bands and of the rebrightening in the UVOT data during the decay of the second outburst. The dash-dotted lines indicate the times of the highest observed X-ray flux and of the X-ray peak of the third and fourth outburst.}
\label{Fig:lc_uvot}
\end{figure}

UVOT data are available in the V, U, UVW1, UVW2, and UVM2 bands. Not all bands are covered in each observation. Light curves in the different bands together with the \swift/XRT light curve are shown in Fig.~\ref{Fig:lc_uvot}. The beginning of the outburst is well covered in the UVW2 band. In this band, as well as in the other two UV bands, the first outburst appears to be much more peaked than in the X-ray data and the first plateau seen in the X-ray band is not present in the UV data. The coverage of the beginning of the outburst in the V band is spare, while in the U band the detector saturates in many observations when the source is at its brightest stage, which results in a constant magnitude with a large error bar. Comparing the UV bands to the X-ray data, we also notice that the second outburst, which is brighter than the first one in the X-ray data, does not reach the magnitudes of the first outburst in the UV bands. We also notice that during the decay of the second outburst a rebrightening occurs in all UVOT bands around day 214. This feature is most prominent in the V band, and it is not seen in the corresponding X-ray data. The rebrightening takes place when the source evolves along the lower horizontal branch in the HID (Fig.~\ref{Fig:HID}) in the LHS at \swift/XRT HRs between 0.14 and 0.22.  

\begin{figure}
\centering
\resizebox{\hsize}{!}{\includegraphics[clip,angle=0]{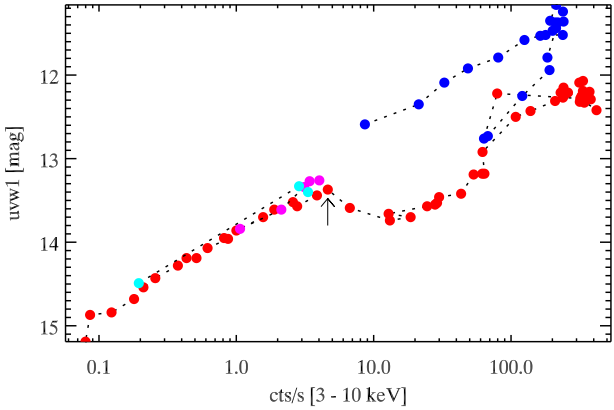}}
\caption{Correlation between the \swift/XRT count rate in the hard (3 -- 10 keV) band and the magnitude in the near ultraviolet (UVW1) band. Each data point represents one observation. Observations of the four outbursts are indicated by different colours, following the colour scheme of Fig.~\ref{Fig:LC}. The arrow indicates the observation in which the rebrightening seen in the UVOT bands peaks.}
\label{Fig:uvx}
\end{figure}

Investigating the correlation between fluxes in the near ultraviolet band (UVW1) and the \swift/XRT hard X-ray band (see Fig.~\ref{Fig:uvx}) we find a linear correlation of these two quantities at the beginning of the first outburst, before the first plateau in the X-rays. The same correlation is found for data points corresponding to observations taken after the peak of the rebrightening in the UVOT bands. The data points of the third and fourth outburst also follow this correlation rather closely. These findings can be interpreted that it is the same process that creates the UV emission at the beginning of the outburst and after the peak of the rebrightening, while the UV emission in between is due to another process. The rebrightening in the UVOT bands indicates a change in the accretion morphology \eg\ in the structures of the accretion disk or jet.

\subsection{X-ray timing properties}
\label{Sec:time}
\subsubsection{\swift/XRT}
\label{Sec:time_sw}
\begin{figure*}
\resizebox{\hsize}{!}{\includegraphics[clip,angle=0]{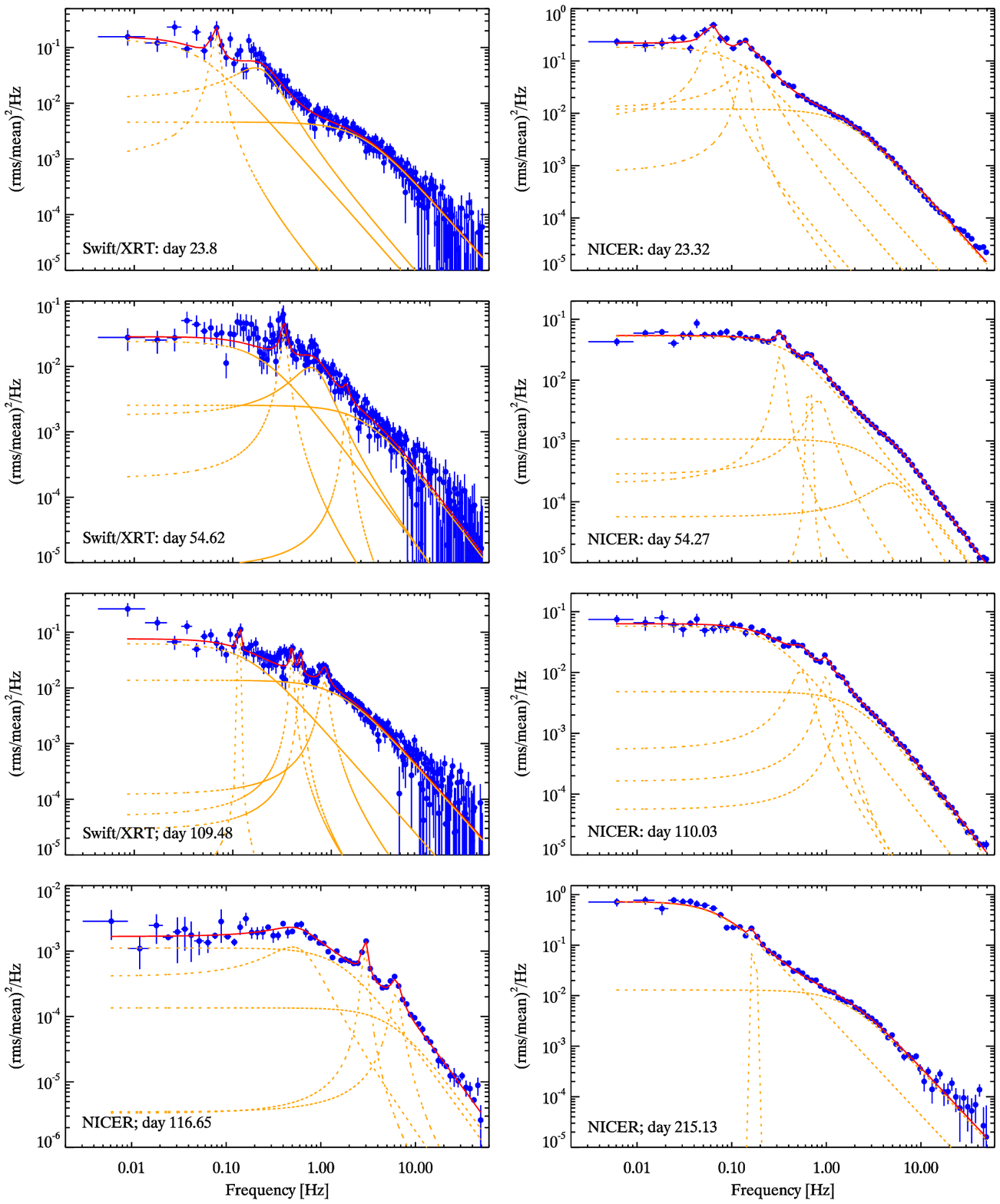}}
\caption{Examples of PDS for three \swift/XRT observations, taken during brightening of the outburst. PDS of corresponding NICER observations are also shown. In addition, two NICER PDS obtained on day 116 and day 215 are shown.}
\label{Fig:pds}
\end{figure*}

\begin{figure}
\resizebox{\hsize}{!}{\includegraphics[clip,angle=0]{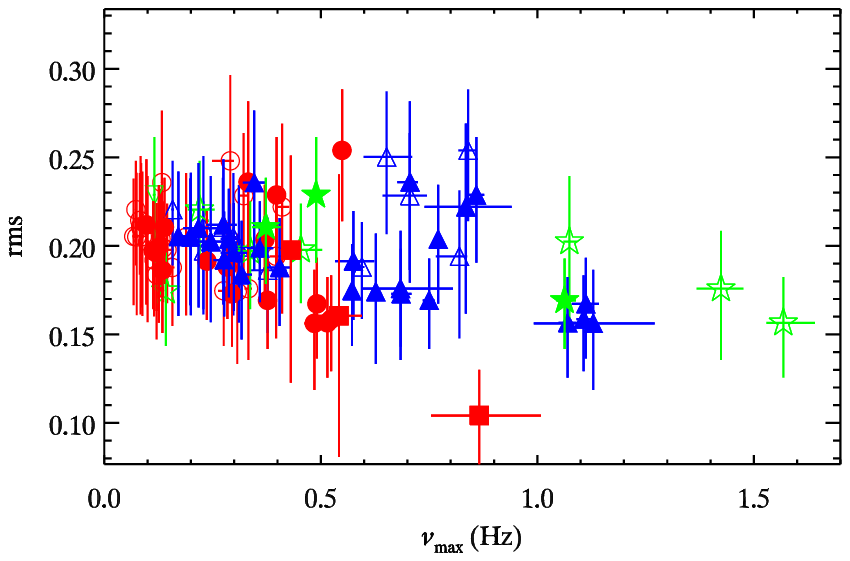}}\\ 
\resizebox{\hsize}{!}{\includegraphics[clip,angle=0]{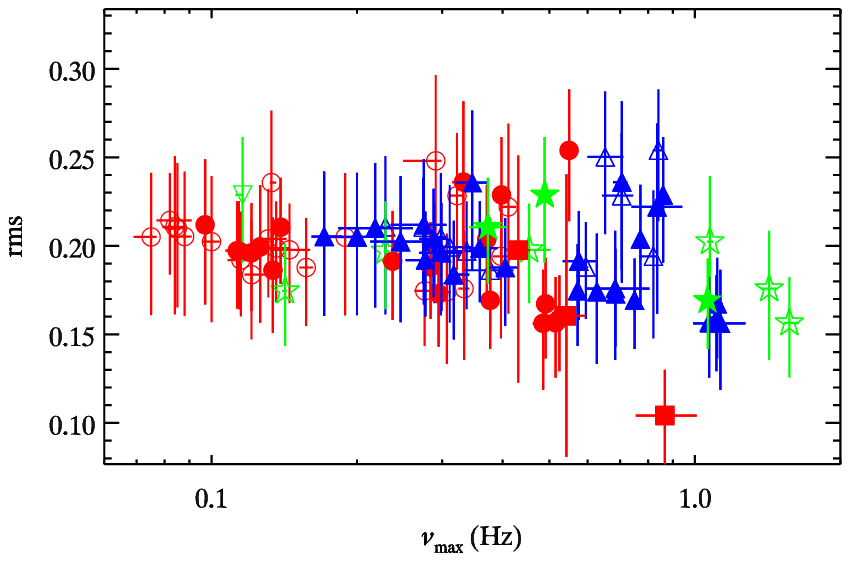}}
\caption{Correlation of the total fractional rms variability with the characteristic QPO frequency, derived from Swift/XRT data. Circles indicate observations taken during the first outburst and the rise of the second outburst, while squares indicate observations taken during the decay of the second outburst (open symbols indicate QPOs with a detection significance below $3\sigma$). For observations in which more than one QPO is detected (only the case during outburst rise) up-pointing triangles, stars, and down-pointing triangles indicate QPOs at higher frequencies (including upper harmonics). In the lower panel the x-axis is logarithmic to display the data points at lower frequencies more clearly. }
\label{Fig:fchar_rms}
\end{figure}

The PDS of the observations taken until day 22.4 can be well described by two BLN components. From day 23.8 onwards a QPO is present and it can be detected in most observations of the first plateau and during the first decay and at the beginning of the second rise. Observations where the PDS shows a QPO have a \swift/XRT count rate between $\sim$400 and 1280 cts/s and a total fractional rms between 15.6 and 25.4\%. The characteristic frequency ranges from 0.08 to 0.86 Hz, which only covers the low frequency end of the range in which QPOs are typically observed in black hole XRBs. For most observations the Q factor is between $\sim$3 and 10. Details on individual observations can be found in Tables~\ref{Tab:pds1} and \ref{Tab:pds2} and examples of PDS are shown in Fig.~\ref{Fig:pds}. The PDS of observations taken during the HSS (total fractional rms $\sim$2\%) are dominated by power law noise. The first three observations where the rms is above 10\% show again QPOs with a characteristic frequency between 0.43 and 0.87 Hz. The observed Q factors are low ($<3$). The PDS of the remaining observations can be described by one or two BLN components and no QPO is detected. The correlation between total fractional rms and characteristic frequency is shown in Fig.~\ref{Fig:fchar_rms}. The PDS of the observations during the third outburst can be well fitted by two BLN components. In the observation taken on day 374.12 a QPO with a characteristic frequency of $\sim 6.19$ Hz and a Q factor of 18.4 is detected at  $2.97\sigma$.

\begin{figure}
\resizebox{\hsize}{!}{\includegraphics[clip,angle=0]{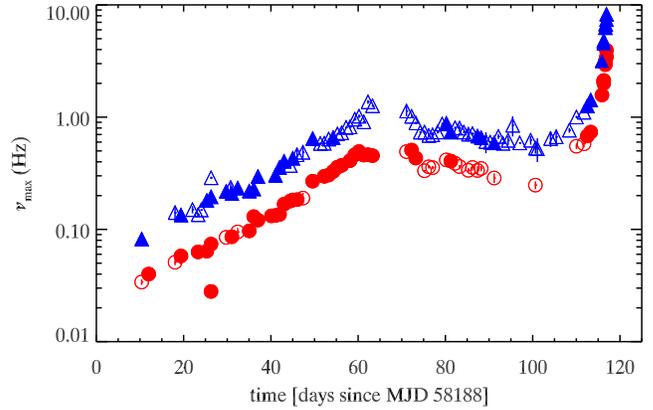}}
\caption{Evolution of the characteristic frequency of the fundamental QPO (red circles) and its upper-harmonic (blue triangles) derived from NICER data until day 116. Open symbols indicate QPOs with a detection significance below $3\sigma$.}
\label{Fig:qpo_evolv}
\end{figure}

\begin{figure}
\resizebox{\hsize}{!}{\includegraphics[clip,angle=0]{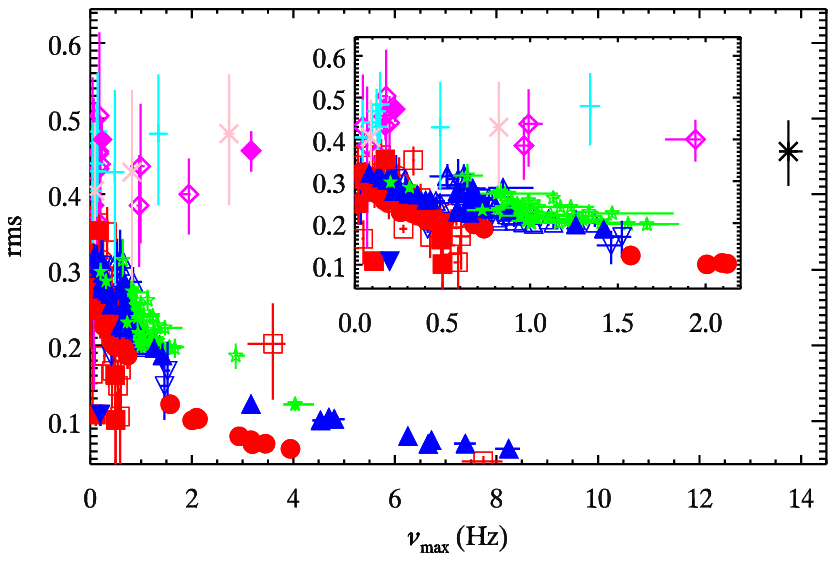}}\\ 
\resizebox{\hsize}{!}{\includegraphics[clip,angle=0]{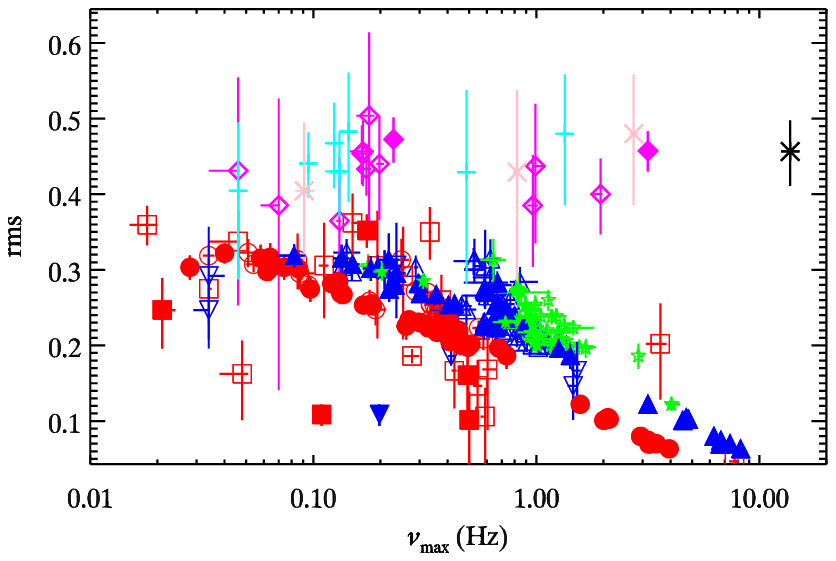}}
\caption{Correlation of the total fractional rms variability with the characteristic QPO frequency, derived from NICER data. Circles indicate observations taken during the first outburst and the rise of the second outburst, while squares indicate observations taken during  the decay of the second outburst (open symbols indicate QPOs with a detection significance below $3\sigma$). For observations in which more than one QPO is detected up-pointing (down-pointing) triangles indicate QPOs at the higher frequency (including upper harmonics) in observations taken during outburst rise (decay). In some observations taken during outburst rise (before day 117) two Lorentizans are need to fit the QPO at higher frequency. If this is the case, green stars indicate this additional feature. Observations taken during the third outburst (after day 330) are indicated using a diamond shape. The black asterisk indicates the QPO at 13.7 Hz found in the observation at the peak of the third outburst. The QPOs and upper harmonics observed during the fourth outburst are indicted by cyan cross points and pink x-shape points, respectively. The inset displays data points at frequencies below 2 Hz more clearly, while in the lower panel the x-axis is logarithmic to display the data points at even lower frequencies ($<0.5$ Hz) more clearly.
}
\label{Fig:fchar_rms_ni}
\end{figure}

\subsubsection{NICER}
Like for the \swift/XRT data, the PDS of the NICER observations taken until day 5.2 can be well described by two BLN components. In most of the NICER observations taken after day 10 and until day 116 two QPOs are detected, with the characteristic frequency increasing from 0.03 to 1.57 Hz and from 0.08 to 3.17 Hz, respectively, although in some observations these features are detected at less than three sigma. The Q values lie between 2 and 10 for most observations. In some observations the lower detection significance is related to the fact that two Lorentzians, a broader and a narrow one, are required to obtain a proper fit of the QPO. In other observations a rather sharp, edge like feature is visible, which gives a high Q factor ($>10$) if fitted with a Lorentzian. Between day 35 and 116 most observations show an additional peaked noise component, with a characteristic frequency between 3 and 8 Hz. Details on individual observations can be found in Tables~\ref{Tab:pds1ni} and \ref{Tab:qpo_ni} and examples of PDS are shown in Fig.~\ref{Fig:pds}. From Fig.~\ref{Fig:qpo_evolv}, which shows the evolution of the characteristic frequency up until day 116, it seems that for most observations taken between day 92 and 109 only the QPO at higher frequency is detectable. The data point at lower frequencies on day 26.28 indicates that on this day the sub-harmonic was detected by chance (together with the QPO).

On day 116 two QPOs are detected, with the characteristic frequency increasing from 2 to 4 Hz and from 4 to 8 Hz, respectively. At the beginning of this observation the higher frequency QPO, which has a lower Q factor, is detected at a higher significance than the lower frequency QPO. Then a third QPO with a characteristic frequency around 0.6 -- 0.8 Hz appears. After this third QPO has disappeared again, the detection significance of the other two QPOs differs much less. The evolution of these QPOs is shown in Fig.~\ref{Fig:qpo_evolv} and details are given in Table~\ref{Tab:qpo_ni}.  

In the observations taken after 2018 September 25th, QPOs with $>2.9\sigma$ and $Q>2$ are detected on days 199, 215 ,233, 241, 246 and have characteristic frequencies of 0.11, 0.17, 0.02, 0.49, 0.50 Hz, respectively. There are a few more QPOs with high Q values and characteristic frequencies between 0.6 and 0.03 Hz, that are detected around $2\sigma$. These QPOs are also included in Table~\ref{Tab:qpo_ni}.

During the third and fourth outburst (observations taken in 2019) we find some more QPOs with high Q values that are detected around $2\sigma$. Two QPOs are detected with more than 3$\sigma$ on days 375 and 380 and have characteristic frequencies of 0.23 and 3.17 Hz, respectively. At the brightest point of the third outburst at day 372 we find a QPO at a characteristic frequency of 13.75 Hz that is detected at 2.8$\sigma$. These QPOs are also included in Table~\ref{Tab:qpo_ni}.

The correlation between total fractional rms and characteristic frequency is shown in Fig.~\ref{Fig:fchar_rms_ni}. Most QPOs detected at $>3\sigma$ follow an anti correlation between rms and frequency. This anti correlation holds from the lowest frequency of 0.03 Hz to the highest at 8 Hz, and suggests that the QPOs are of type-C. The QPOs observed between day 92 and 109, where only the QPO at higher frequency is detected, seem to follow a steeper anti-correlation than the remaining QPOs. It also seems that the QPOs observed during the third and fourth outburst (most of them only detected $<3\sigma$) follow a flat correlation at $\sim43$\% rms.

\begin{figure*}
\resizebox{\hsize}{!}{\includegraphics[clip,angle=0]{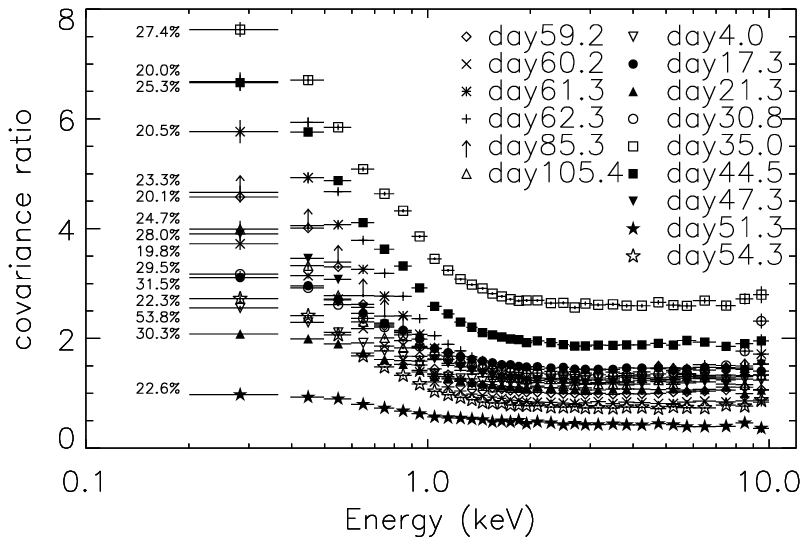}}
\caption{Covariance ratios of a sample of observations taken during the low-hard state with NICER. The small percentage numbers indicate the fractional rms values of the individual observations. The steepness of the increase towards lower energies neither correlates with the amount of rms variability nor with the outburst evolution.}
\label{Fig:cov_ratio}
\end{figure*}

\begin{figure}
\resizebox{\hsize}{!}{\includegraphics[clip,angle=0]{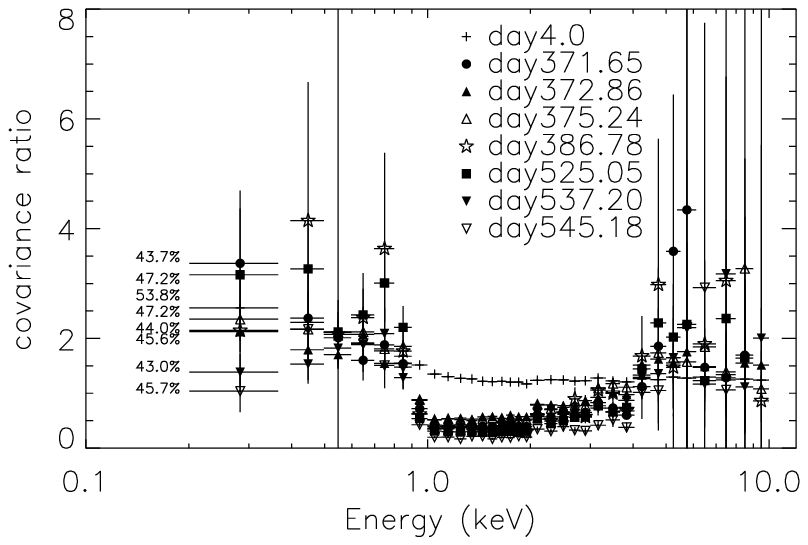}}
\caption{Covariance ratios of a sample of observations taken during the third and fourth outburst with NICER. The small percentage numbers indicate the fractional rms values of the individual observations. For comparison the covariance ratios for the observation on day 4 are given.}
\label{Fig:cov_ratio2}
\end{figure}

We derive covariance spectra on long (5 s; 20 bins) and short (0.01 s; 50 bins) time scales \citep{2009MNRAS.397..666W,2015MNRAS.452.3666S}. The covariance ratios, obtained by dividing  spectra on long time scales by those on short time scales, for a sample of observations, are shown in Fig.~\ref{Fig:cov_ratio}, and show an increase towards lower energies, as observed \eg\ in \gx339\ and Swift\,J1753.5-0125 \citep{2009MNRAS.397..666W,2015MNRAS.452.3666S}. Although the steepness of the increase changes over time it is present throughout the hard state over more than 100 days. We also find that the steepness of the increase is not correlated with the amount of rms variability and that it does not show a monotonic evolution along the outburst (see Fig.~\ref{Fig:cov_ratio}). We also derived covariance spectra and ratios for observations of the third outburst (Fig.~\ref{Fig:cov_ratio2}). Due to the coverage of this outburst by NICER data, we only have one observation before the outburst peak (and several observations after the peak). Considering only covariance ratios above 4 keV and below 0.8 keV, the ratios appear to be much flatter than the correlations observed during the first and second outburst. A comparison of the covariance ratios for the different outbursts in the 0.8 to 4 keV range shows that the covariance ratios of the third outburst show a decrease towards lower energies in this energy range. The covariance ratio of the fourth outburst that are obtained in a similar count rate range as those of the third outburst show the same shape as those of the third outburst. Although earlier phases of this outburst are covered compared to the coverage of the third outburst, we cannot determine the shape of the covariance ratios in these observations due to the low count rate.

\section[]{Discussion}
\label{Sec:dis}
We investigate the \swift/XRT and NICER monitoring observations of \max18\ taken in 2018 and 2019. The HID derived from the \swift/XRT data show the typical q-shape or turtle-head pattern observed in many black hole X-ray binary outbursts. The light curve reveals that \max18\ underwent four outbursts, with the source remaining in the hard state during its first outburst, while the rise of the second outburst corresponds with the transition to the HSS. These two outbursts are followed by two weaker outbursts during which \max18\ remains again in the hard state, so-called ``failed" outbursts. The first two outbursts can be regarded as a double outburst and a similar outburst of \gx339\ has been observed in 2004 \citep{2007ApJ...657..400J,2014MNRAS.442.1767P}. In case of \gx339\ the outburst lasted about twice as long as the double outburst of \max18\ reported here. Comparing the UV to the X-ray light curves shows that the first outburst, which is brighter than the second outburst in the UV bands, is much more peaked in the UV bands than in the X-ray band, and that the plateau seen in the X-ray data is not present in the UV data. During the decay of the second outburst a rebrightening is visible in the UVOT data that has no corresponding feature in the X-ray light curves.  

For many observations during the first outburst and the rise of the second outburst QPOs have been detected in the \swift/XRT and NICER data. For most of these observations the characteristic frequencies of the QPOs detected at about the same time in both data sets agree with each other. Due to the much better signal-to-noise ratio in the NICER PDS, QPOs at lower frequencies and weaker features (like upper harmonics or additional peaked noise components) can be detected in these data, which are not detectable in \swift/XRT PDS. For the QPOs observed during outburst rise (up until day 116), the frequency range in which the QPOs are observed, the Q factors and the anti-correlation between total fractional rms and characteristic frequency suggest these QPOs to be type-C QPOs, consistent with the detection of these QPOs during the hard state. The QPOs detected during the first plateau, the first decay and at the beginning of the second rise are detected at rather low characteristic frequencies below 1 Hz. Only in the last two NICER observations taken before the source entered the soft state, QPOs with a characteristic frequency above 1 Hz are detected and the characteristic frequency increases from 1.57 to 3.94 Hz within 24.8 hrs, while the overall total fractional rms decreases from $10.4^{+1.2}_{-1.1}$ to $6.3^{+1.1}_{-1.0}$\%.  In the \swift/XRT observations taken about 19 hrs after the last NICER observation the rms has decreased to $2^{+3}_{-2}$\% and \max18\ has entered the soft state, in which it stayed for about 81 days. On day 199.4 the rms has increased to $10.9^{+1.5}_{-1.3}$\% and two QPOs with characteristic frequencies of 0.11 and 0.20 Hz are detected. On day 202 and 203 QPOs with characteristic frequencies of about 0.5 and 0.4 Hz, respectively, are detected in both \swift/XRT and NICER data. A few more QPOs are detected in the NICER data during the decay of the second outburst and during the two ``failed'' outbursts. 

At the brightest point of the third outburst at day 372 we find a QPO at $\sim13.75$ Hz. It is worth noting that a QPO at $\sim21$ Hz has been observed in the initial low-hard state of XTE\,J1752--223 \citep{2010MNRAS.404L..94M}. 

We would like to point out that during the first outburst and during most of the times of the second, third and fourth outburst, the characteristic frequency of the QPO remains below 1 Hz, while in other black hole X-ray binaries that normally show single-peak outbursts (including so-called failed outbursts), the characteristic frequency of type-C QPOs increases up to 10 Hz \citep{2013MNRAS.429.2655S}. The observation of type-C QPOs below 1 Hz during large parts of an outburst of a black hole X-ray binary is exceptional. A comparison of the QPOs and their properties observed here to those detected in the 2004 outburst of \gx339, shows that during the 2004 outburst only one QPO with a characteristic frequency below 1 Hz was detected \citep{2011MNRAS.418.2292M}. QPO-like features at rather low characteristic frequencies (0.03 and 0.05 Hz) have also been observed in the initial low-hard state observations of XTE\,J1752--223 \citep{2010MNRAS.404L..94M}, but during its evolution the source showed QPOs at frequencies above 1 Hz in many observations \citep{2010ApJ...723.1817S}. QPOs at frequencies between 0.1 and 1 Hz have also be detected in Cyg\,X--1 \citep{2003A&A...407.1039P}. The QPOs below 1 Hz do not show a correlation with the source flux, as has been observed for example in \citet{2014ApJ...789..100S}. During the first plateau we observe an increase in the characteristic frequency from $\sim 0.04$ to $\sim0.3$ Hz, while the source luminosity is slightly decreasing. 

In case of the 2004 double outburst of \gx339\ only in one observation during the decay of the second outburst a QPO was detected. For \max18\ we found five QPOs in the NICER data of the decay of the second outburst. So it seems that double outbursts show QPOs only in few observations during  the decay of the second outburst compared to the higher numbers of QPOs observed during the decay of single-peak outbursts of other black hole X-ray binaries \citep{2013MNRAS.429.2655S}. However, one needs to be cautious when comparing the amount of QPOs detected in individual outbursts and in different parts of an outburst, as the number of QPOs found not only depends on the properties of the system, but also on the amount of observations that are available for each part of an outburst.

It is also remarkable that the increase in QPO frequency and the transition to the soft state take place in less than two days. This is an exceptional fast state transition (see Table~16 of \citet{2016ApJS..222...15T} that lists durations of state transitions for transient Galactic BHBs). 

The detection of a disk wind in the optical during the hard state \citep{2019ApJ...879L...4M} together with the fast transition from the hard to the soft state, seems to indicate the presence of a standard accretion disk in the hard state of \max18. However, the fact that the characteristic frequencies of the QPOs remains below 1 Hz for most of the outburst seems to imply that the source remains in a configuration close to the one typically observed at the beginning of an outburst, when the accretion efficiency is still low and the disk is truncated far away from the black hole. In the Lense-Thirring precession model \citep{2009MNRAS.397L.101I} low QPO frequencies correspond to an accretion disk truncated at several tens of gravitational radius from the central black hole. Hence the observed low QPO frequencies imply an accretion disk truncated far away from the black hole and not much evolution of the truncation radius during the outburst. This is consistent with the results of \citet{2019Natur.565..198K} that the truncation radius of the accretion disk does not show much evolution during the hard state based on X-ray reverberation lags. An explanation why the source remains in a state of low accretion efficiency can be the presence of the disk wind in the hard state that has not been observed in any other black hole X-ray binary before, and that hampered the formation of a stable accretion regime.  

We also investigate \swift/UVOT light curves. In the UV light curves the first outburst is brighter than the second one, contrary to what has been observed in the X-rays, and it is much more peaked than in the X-rays. The source of UV emission in the hard state of BHBs can be the jet, the cool disk, the hot advection-dominated accretion flow or X-ray reprocessing in the accretion disk \citep{2003A&A...397..645M,2005ApJ...620..905Y, 2007ApJ...666.1129R,2009MNRAS.398.1638M}. As the shape of the light curves differs between the X-ray and UV bands, viscous dissipation in the disk does not seem to be the primarily cause of the UV emission. Jet quenching, which has been suggested as cause of a similar feature observed in \swift\ data of the 2010 outburst of \gx339\ \citep{2012MNRAS.427L..11Y}, also does not seem to be a good explanation here, as \max18\ remains in the hard state for about another 100 days, before the transition to the soft state takes place.     

Furthermore, the \swift/UVOT light curves show a rebrightening that peaked about 15 days after the source left the HSS. The X-ray light curves do not show a corresponding feature and keep on decaying. The rebrightening was accompanied by an increase in the X-ray hardness ratio, which indicates a change in the accretion morphology. This change in the accretion morphology is further supported by the finding that a linear correlation between the hard X-ray count rate and the near UV flux similar to the one at the beginning of the outburst is observed after the peak of the rebrightening feature. A similar UV rebrightening with continued decay in the X-rays and an increase in the X-ray hardness ratio was observed in the 2012 outburst of SWIFT\,J1910.2--0546 \citep{2014ApJ...784..122D}. The rebrighetning in the UVOT bands is reminiscent of the secondary nIR/optical maxima seen in other BHBs that have been ascribed to the revival of a jet \citep[e.g.][]{2012ApJ...753...55D,2013ApJ...779...95K}. Alternatively, the rebrightening may be interpreted as the recovery of a hot inner flow \citep{2011MNRAS.414.3330V,2013MNRAS.430.3196V}.

The covariance ratios derived from NICER data of the first and second outburst show an increase towards lower energies. In previous studies, we related the increase in covariance ratio to outbursts making a transition to the high-soft state, while for outbursts that remain in the hard states, we observed a flat or decreasing covariance ratio \citep{2016MNRAS.459.4038S,2016MNRAS.460.1946S}. Thus the increase in the covariance ratios of \max18\ indicated that the source should make a transition to the soft state, which in the end took place, after \max18\ stayed in the hard state at rather constant luminosity for about 116 days. A similar shape of the covariance ratio has also been found from \xmm\ observations taken during the first outburst of the 2004 double-outburst of \gx339\ \citep{2019AN....340..314S}.  The steepness of the increase is not correlated with the amount of rms variability and it does not show a monotonic evolution along the outburst, contrary to what has been observed during the 2015 outburst decay of \gx339\ \citep{2017ApJ...844....8S}. This finding gives further evidence that there is no correlation between the shape of the covariance ratio and the amount of fractional rms variability, as reported in \citet{2019AN....340..314S}.

The overall shape of the covariance ratios of the third and fourth, hard state only outbursts clearly differs from the shape seen in the first two outbursts. Between 0.8 and 4 keV the covariance ratios decay towards lower energies. There seems to be a slight increase to even lower energies and the shape is different from the ones reported in \citet{2016MNRAS.459.4038S} and \citet{2016MNRAS.460.1946S}. This difference may be related to the phase of the outburst during which the observations have been taken. Here, covariance ratios are obtained from NICER observations of \max18\ shortly before the peak of its third and fourth outburst and the outburst decays are well covered. Contrary, the observations studied in \citet{2016MNRAS.459.4038S} and \citet{2016MNRAS.460.1946S} have been taken during early outburst rise of the hard state only outbursts. Furthermore, the two hard state only outbursts of \max18\ follow a bright double outburst, while the outbursts studied in \citet{2016MNRAS.459.4038S} and \citet{2016MNRAS.460.1946S} are single hard state only outbursts that are separated by much longer time span from any previous outburst. Hence, the accretion geometry from the hard state only outbursts studied here may be different from the one in the outbursts studied in \citet{2016MNRAS.459.4038S} and \citet{2016MNRAS.460.1946S}. More observations of other hard state only outburst will be needed to study the evolution of the covariance ratios during this kind of outburst and to further investigate differences between full and hard state only outbursts.


\acknowledgments
We like to thank the anonymous reviewer for his/her thoughtful comments that have been helpful in improving the presentation of our results.
This project is supported by the Ministry of Science and Technology of
the Republic of China (Taiwan) through grants 105-2119-M-007-028-MY3 and 107-2811-M-007-029.
This research has made use of data obtained through the High Energy Astrophysics Science Archive Research Center Online Service, provided by the NASA/Goddard Space Flight Center. This research has made use of the General High-energy Aperiodic Timing Software (GHATS) package developed by T.\ M.\ Belloni at INAF Osservatorio Astronomico di Brera. This work makes use of software tools provided by Simon Vaughan.

{\it Facilities:} \facility{\swift, NICER}.
\software{HEAsoft, Isis, GHATS}
\clearpage

\bibliographystyle{aasjournal}
\bibliography{}

\appendix
\begin{table*}
\caption{Parameters of the BLN components of the PDS derived from \swift/XRT data}
\begin{center}
\begin{tabular}{rrrrr}
\hline\noalign{\smallskip}
 \multicolumn{1}{c}{day} & \multicolumn{1}{c}{$\nu_{\mr{max; BLN1}}$} & \multicolumn{1}{c}{rms$_{\mr{BLN1}}$} & \multicolumn{1}{c}{$\nu_{\mr{max; BLN2}}$} & \multicolumn{1}{c}{rms$_{\mr{BLN2}}$} \\ 
\hline\noalign{\smallskip}
2.74  &  $0.974_{-0.091}^{+0.121} $  &  $0.322_{-0.009}^{+0.008}  $  &  $0.020_{-0.005}^{+0.004}  $  &  $0.408_{-0.031}^{+0.039}  $  \\ 
\smallskip
3.84  &  $1.404_{-0.116}^{+0.199} $  &  $0.278_{-0.020}^{+0.013}  $  &  $0.266_{-0.059}^{+0.075}  $  &  $0.239_{-0.016}^{+0.020}  $  \\ 
\smallskip
3.87  &  $1.116_{-0.075}^{+0.075} $  &  $0.326_{-0.006}^{+0.006}  $  &  $0.025_{-0.004}^{+0.007}  $  &  $0.358_{-0.028}^{+0.026}  $  \\ 
\smallskip
4.08  &  $1.060_{-0.064}^{+0.076} $  &  $0.337_{-0.006}^{+0.005}  $  &  $0.024_{-0.004}^{+0.008}  $  &  $0.374_{-0.028}^{+0.024}  $  \\ 
\smallskip
5.13  &  $1.355_{-0.078}^{+0.075} $  &  $0.282_{-0.005}^{+0.005}  $  &  $0.028_{-0.004}^{+0.006}  $  &  $0.334_{-0.024}^{+0.019}  $  \\ 
\smallskip
6.13  &  $1.180_{-0.068}^{+0.072} $  &  $0.260_{-0.005}^{+0.004}  $  &  $0.025_{-0.003}^{+0.005}  $  &  $0.301_{-0.021}^{+0.017}  $  \\ 
\smallskip
7.13  &  $1.163_{-0.057}^{+0.054} $  &  $0.216_{-0.003}^{+0.004}  $  &  $0.021_{-0.004}^{+0.004}  $  &  $0.245_{-0.017}^{+0.019}  $  \\
\end{tabular} 
\end{center}
\label{Tab:pds1}
Notes:\\ 
rms: root mean square; $\nu_{\mr{max}}$: characteristic frequency; BLN: band limited noise
\end{table*}

\addtocounter{table}{-1}
\begin{table*}
\caption{continued}
\begin{center}
\begin{tabular}{rrrrr}
\hline\noalign{\smallskip}
 \multicolumn{1}{c}{day} & \multicolumn{1}{c}{$\nu_{\mr{max; BLN1}}$} & \multicolumn{1}{c}{rms$_{\mr{BLN1}}$} & \multicolumn{1}{c}{$\nu_{\mr{max; BLN2}}$} & \multicolumn{1}{c}{rms$_{\mr{BLN2}}$} \\ 
\hline\noalign{\smallskip}
36.69  &  $1.918_{-0.123}^{+0.139} $  &  $0.113_{-0.004}^{+0.003}  $  &  $0.107_{-0.027}^{+0.030}  $  &  $0.110_{-0.016}^{+0.016}  $  \\ 
\smallskip
36.75  &  $2.086_{-0.146}^{+0.176} $  &  $0.113_{-0.005}^{+0.003}  $  &  $0.117_{-0.034}^{+0.060}  $  &  $0.103_{-0.019}^{+0.028}  $  \\ 
\smallskip
36.83  &  $1.924_{-0.157}^{+0.172} $  &  $0.101_{-0.003}^{+0.003}  $  &  $0.104_{-0.029}^{+0.021}  $  &  $0.117_{-0.018}^{+0.009}  $  \\ 
\smallskip
36.89  &  $2.145_{-0.146}^{+0.173} $  &  $0.121_{-0.004}^{+0.004}  $  &  $0.117_{-0.047}^{+0.037}  $  &  $0.113_{-0.027}^{+0.015}  $  \\ 
\smallskip
36.96  &  $2.111_{-0.245}^{+0.243} $  &  $0.103_{-0.005}^{+0.005}  $  &  $0.207_{-0.023}^{+0.025}  $  &  $0.141_{-0.005}^{+0.005}  $  \\ 
\smallskip
37.48  &  $2.152_{-0.157}^{+0.182} $  &  $0.099_{-0.003}^{+0.003}  $  &  $0.222_{-0.014}^{+0.015}  $  &  $0.149_{-0.004}^{+0.004}  $  \\ 
\smallskip
37.76  &  $1.946_{-0.167}^{+0.211} $  &  $0.107_{-0.006}^{+0.004}  $  &  $0.124_{-0.030}^{+0.045}  $  &  $0.112_{-0.015}^{+0.019}  $  \\
\end{tabular} 
\end{center}
\label{Tab:pds1}
Notes:\\ 
rms: root mean square; $\nu_{\mr{max}}$: characteristic frequency; BLN: band limited noise
\end{table*}

\addtocounter{table}{-1}
\begin{table*}
\caption{continued}
\begin{center}
\begin{tabular}{rrrrr}
\hline\noalign{\smallskip}
 \multicolumn{1}{c}{day} & \multicolumn{1}{c}{$\nu_{\mr{max; BLN1}}$} & \multicolumn{1}{c}{rms$_{\mr{BLN1}}$} & \multicolumn{1}{c}{$\nu_{\mr{max; BLN2}}$} & \multicolumn{1}{c}{rms$_{\mr{BLN2}}$} \\ 
\hline\noalign{\smallskip}
374.12  &$0.244_{-0.046}^{+0.064}$ & $  0.283_{-0.020}^{+0.019}$ & $  0.010_{-0.005}^{+0.005}$ & $  0.330_{-0.042}^{+0.057}$ \\
\smallskip
375.12  &$0.011_{-0.004}^{+0.006}$ & $  0.286_{-0.034}^{+0.036}$ & $  0.364_{-0.077}^{+0.181}$ & $  0.248_{-0.016}^{+0.015}$ \\
\smallskip
376.26  &$0.009_{-0.003}^{+0.003}$ & $  0.370_{-0.045}^{+0.061}$ & $  0.691_{-0.149}^{+0.203}$ & $  0.282_{-0.017}^{+0.017}$ \\
\smallskip
377.18  &$0.016_{-0.004}^{+0.006}$ & $  0.325_{-0.032}^{+0.028}$ & $  0.615_{-0.226}^{+0.423}$ & $  0.200_{-0.028}^{+0.024}$ \\
\smallskip
378.28  &$0.023_{-0.004}^{+0.006}$ & $  0.373_{-0.023}^{+0.023}$ & $  1.214_{-0.194}^{+0.391}$ & $  0.287_{-0.014}^{+0.017}$ \\
\smallskip
379.21  &$0.034_{-0.009}^{+0.018}$ & $  0.273_{-0.019}^{+0.019}$ & $  4.377_{-0.490}^{+0.614}$ & $  0.449_{-0.018}^{+0.019}$ \\
\smallskip
383.03  &$0.008_{-0.004}^{+0.005}$ & $  0.442_{-0.083}^{+5.682}$ & $  6.906_{-1.420}^{+1.850}$ & $  0.496_{-0.043}^{+0.044}$ \\
\smallskip
384.15  &$0.032_{-0.010}^{+0.012}$ & $  0.241_{-0.020}^{+0.020}$ & $  8.398_{-1.050}^{+1.191}$ & $  0.614_{-0.031}^{+0.031}$ \\
\smallskip
530.24& $0.0257_{-0.0115}^{+0.0128}$ & $	0.3176_{-0.0303}^{+0.0406}$  &$0.683_{-0.2601}^{+0.3246}$ & $	0.2045_{-0.0331}^{+0.0288}$ \\
\smallskip
531.11 & $0.0104_{-0.0033}^{+0.0055}$ & $	0.301_{-0.0437}^{+0.0432}$ &$3.4271_{-1.6229}^{+2.4425}$ & $	0.2927_{-0.0423}^{+0.0479}$ \\
\smallskip
536.02 & $0.0119_{-0.0062}^{+0.0066}$ & $	0.271_{-0.0459}^{+0.0373}$ &$0.1751_{-0.0764}^{+0.2161}$ & $	0.2036_{-0.0397}^{+0.0343}$ \\
\smallskip
545.05 &$0.012_{-0.0042}^{+0.005}$ & $	0.3088_{-0.0397}^{+0.0428}$ & 									& \\
\hline\noalign{\smallskip}
\end{tabular} 
\end{center}
\label{Tab:pds1}
Notes:\\ 
rms: root mean square; $\nu_{\mr{max}}$: characteristic frequency; BLN: band limited noise
\end{table*}

\begin{table*}
\caption{Parameters of the QPOs of the PDS derived from \swift/XRT}
\begin{center}
\begin{tabular}{rrrrrr}
\hline\noalign{\smallskip}
 \multicolumn{1}{c}{day} & \multicolumn{1}{c}{$\nu_{\mr{0; QPO1}}$} & \multicolumn{1}{c}{$\Delta_{\mr{QPO1}}$} & \multicolumn{1}{c}{rms$_{\mr{QPO1}}$} & \multicolumn{1}{c}{Q$_{\mr{QPO1}}$}  & \multicolumn{1}{c}{$\sigma_{\mr{QPO1}}$}\\ 
\hline\noalign{\smallskip}
23.80 & $0.0675_{-0.0023}^{+0.0029}$  &$	0.0049_{-0.0035}^{+0.0043}$  &$	0.0548_{-0.0132}^{+0.0114}$   &6.89	&2.08\\
\smallskip
24.24 & $0.064_{-0.001}^{+0.004}$  &$<	0.006$  &$	0.043_{-0.012}^{+0.014}$   &5.74	&1.81\\
\smallskip
26.88 & $0.0729_{-0.0002}^{+0.0014}$  &$	<0.0028$  &$	0.0411_{-0.0084}^{+0.0071}$   & 13.02	&2.45 \\
\smallskip
28.31 &  $0.075_{-0.002}^{+0.006}$  &$0.002_{-0.002}^{+0.004}$  &$0.042_{-0.012}^{+0.011}$  &23.41  &1.81  \\ 
\smallskip
29.41 &  $0.082_{-0.008}^{+0.004}$  &$0.009_{-0.005}^{+0.010}$  &$0.054_{-0.010}^{+0.010}$  &4.76  &2.71  \\ 
\smallskip
30.50 &  $0.085_{-0.002}^{+0.004}$  &$0.005_{-0.003}^{+0.010}$  &$0.061_{-0.011}^{+0.011}$  &7.97  &2.70  \\ 
\smallskip
32.10 &  $0.087_{-0.003}^{+0.006}$  &$0.008_{-0.004}^{+0.009}$  &$0.064_{-0.011}^{+0.009}$  &5.74  &2.84  \\
\end{tabular} 
\end{center}
\label{Tab:pds2}
Notes:\\ 
rms: root mean square; $\nu_0$: centroid frequency; $\Delta$: half width at half maximum; $\sigma$: significance; QPO: quasiperiodic oscillation
\end{table*}

\addtocounter{table}{-1}
\begin{table*}
\caption{continued}
\begin{center}
\begin{tabular}{rrrrrr}
\hline\noalign{\smallskip}
 \multicolumn{1}{c}{day} & \multicolumn{1}{c}{$\nu_{\mr{0; QPO1}}$} & \multicolumn{1}{c}{$\Delta_{\mr{QPO1}}$} & \multicolumn{1}{c}{rms$_{\mr{QPO1}}$} & \multicolumn{1}{c}{Q$_{\mr{QPO1}}$}  & \multicolumn{1}{c}{$\sigma_{\mr{QPO1}}$}\\ 
\hline\noalign{\smallskip}
108.61&  $0.410_{-0.015}^{+0.013}$  &$0.023_{-0.014}^{+0.016}$  &$0.043_{-0.010}^{+0.010}$  &9.11  &2.10  \\ 
\smallskip
109.48&  $0.397_{-0.006}^{+0.004}$  &$0.014_{-0.007}^{+0.008}$  &$0.042_{-0.007}^{+0.006}$  &14.29  &3.20  \\ 
\smallskip
200.96&  $0.760_{-0.090}^{+0.066}$  &$0.416_{-0.098}^{+0.084}$  &$0.061_{-0.010}^{+0.012}$  &0.91  &3.10  \\ 
\smallskip
202.32&  $0.507_{-0.034}^{+0.039}$  &$0.192_{-0.050}^{+0.064}$  &$0.083_{-0.011}^{+0.011}$  &1.32  &3.72  \\ 
\smallskip
203.46&  $0.425_{-0.020}^{+0.015}$  &$0.071_{-0.025}^{+0.049}$  &$0.070_{-0.010}^{+0.014}$  &2.97  &3.48  \\ 
\smallskip
374.12& $6.190_{-0.000}^{+0.000}$  &$0.169_{-0.075}^{+0.122}$  &$0.140_{-0.024}^{+0.023}$  &18.36 &2.97 \\
\smallskip
531.11& $0.2553_{-0.0103}^{+0.0096}$  &$0.0263_{-0.0098}^{+0.0122}$  &$0.1448_{-0.0213}^{+0.0194}$  &	4.85	&3.40\\
\end{tabular} 
\end{center}
\label{Tab:pds2}
Notes:\\ 
rms: root mean square; $\nu_0$: centroid frequency; $\Delta$: half width at half maximum; $\sigma$: significance; QPO: quasiperiodic oscillation
\end{table*}

\addtocounter{table}{-1}
\begin{table*}
\caption{continued}
\begin{center}
\begin{tabular}{rrrrrr}
\hline\noalign{\smallskip}
 \multicolumn{1}{c}{day} & \multicolumn{1}{c}{$\nu_{\mr{0; QPO2}}$} & \multicolumn{1}{c}{$\Delta_{\mr{QPO2}}$} & \multicolumn{1}{c}{rms$_{\mr{QPO2}}$} & \multicolumn{1}{c}{Q$_{\mr{QPO2}}$}  & \multicolumn{1}{c}{$\sigma_{\mr{QPO2}}$}\\ 
\hline\noalign{\smallskip}
71.42 &  $1.105_{-0.015}^{+0.015}$  &$0.070_{-0.021}^{+0.025}$  &$0.037_{-0.005}^{+0.005}$  &7.92  &3.68  \\ 
\smallskip
79.86 &  $0.819_{-0.017}^{+0.049}$  &$0.050_{-0.024}^{+0.088}$  &$0.036_{-0.008}^{+6.068}$  &8.19  &2.20  \\ 
\smallskip
83.85 &  $0.770_{-0.007}^{+0.009}$  &$0.024_{-0.012}^{+0.017}$  &$0.035_{-0.006}^{+0.006}$  &16.25  &3.04  \\ 
\smallskip
87.83 &  $0.695_{-0.031}^{+0.053}$  &$0.119_{-0.057}^{+0.063}$  &$0.057_{-0.020}^{+0.013}$  &2.91  &1.42  \\ 
\smallskip
92.63 &  $0.838_{-0.017}^{+0.019}$  &$0.055_{-0.027}^{+0.049}$  &$0.039_{-0.008}^{+0.009}$  &7.56  &2.46  \\ 
\smallskip
107.48&  $0.705_{-0.017}^{+0.027}$  &$0.046_{-0.019}^{+0.038}$  &$0.049_{-0.008}^{+0.007}$  &7.60  &3.24  \\ 
\smallskip
108.61&  $0.771_{-0.060}^{+0.044}$  &$0.321_{-0.113}^{+0.122}$  &$0.106_{-0.018}^{+0.017}$  &1.20  &3.00  \\
\end{tabular} 
\end{center}
\label{Tab:pds2}
Notes:\\ 
rms: root mean square; $\nu_0$: centroid frequency; $\Delta$: half width at half maximum; $\sigma$: significance; QPO: quasiperiodic oscillation
\end{table*}

\begin{table*}
\caption{Parameters of the noise components of the PDS derived from NICER data}
\begin{center}
\footnotesize
\begin{tabular}{rrrrrrrrr}
\hline\noalign{\smallskip}
 \multicolumn{1}{c}{day} & \multicolumn{1}{c}{$\nu_{\mr{max; BLN1}}$} & \multicolumn{1}{c}{rms$_{\mr{BLN1}}$} & \multicolumn{1}{c}{$\nu_{\mr{max; BLN2}}$} & \multicolumn{1}{c}{rms$_{\mr{BLN2}}$} & \multicolumn{1}{c}{$\nu_{\mr{max; PN1}}$} & \multicolumn{1}{c}{rms$_{\mr{PN1}}$} & \multicolumn{1}{c}{Q$_{\mr{PN1}}$}  & \multicolumn{1}{c}{$\sigma_{\mr{PN1}}$}\\ 
\hline\noalign{\smallskip}
1.58	&$0.3689_{-0.0479}^{+0.0373} $&$0.334_{-0.0089}^{+0.0056} $&$0.0179_{-0.0019}^{+0.002} $&$0.3995_{-0.0138}^{+0.0132} $&	& & & \\
\smallskip
2.61	&$0.4184_{-0.0309}^{+0.035} $&$0.3502_{-0.0083}^{+0.0046} $&$0.0195_{-0.0023}^{+0.0033} $&$0.4007_{-0.0183}^{+0.0183} $&	& & & \\
\smallskip
3.00	&$0.3816_{-0.0193}^{+0.0136} $&$0.3357_{-0.0041}^{+0.0041} $&$0.0156_{-0.0012}^{+0.0016} $&$0.4285_{-0.013}^{+0.0095} $&	& & & \\
\smallskip
4.03	&$0.4183_{-0.0352}^{+0.0184} $&$0.3404_{-0.0037}^{+0.0065} $&$0.0225_{-0.0028}^{+0.0018} $&$0.3746_{-0.012}^{+0.0125} $&	& & & \\
\smallskip
5.19	&$0.4921_{-0.0275}^{+0.0408} $&$0.2952_{-0.0057}^{+0.007} $&$0.0275_{-0.0022}^{+0.0028} $&$0.35_{-0.011}^{+0.0115} $&	& & & \\
\smallskip
10.39	&$1.3756_{-0.0237}^{+0.0246} $&$0.1926_{-0.0014}^{+0.0018} $&$0.0654_{-0.0039}^{+0.0038} $&$0.2265_{-0.0065}^{+0.0035} $&	& & & \\
\smallskip
11.99	&$0.0675_{-0.0025}^{+0.0026} $&$0.2559_{-0.0036}^{+0.0038} $&$1.5547_{-0.0203}^{+0.0206} $&$0.1839_{-0.0012}^{+0.0012} $&	& & & \\
\end{tabular} 
\end{center}
\label{Tab:pds1ni}
Notes:\\ 
rms: root mean square; $\nu_{\mr{max}}$: characteristic frequency;  $\sigma$: significance; BLN: band limited noise; PN: peaked noise
\end{table*}

\addtocounter{table}{-1}
\begin{table*}
\caption{continued}
\begin{center}
\footnotesize
\begin{tabular}{rrrrrrrrr}
\hline\noalign{\smallskip}
 \multicolumn{1}{c}{day} & \multicolumn{1}{c}{$\nu_{\mr{max; BLN1}}$} & \multicolumn{1}{c}{rms$_{\mr{BLN1}}$} & \multicolumn{1}{c}{$\nu_{\mr{max; BLN2}}$} & \multicolumn{1}{c}{rms$_{\mr{BLN2}}$} & \multicolumn{1}{c}{$\nu_{\mr{max; PN1}}$} & \multicolumn{1}{c}{rms$_{\mr{PN1}}$} & \multicolumn{1}{c}{Q$_{\mr{PN1}}$}  & \multicolumn{1}{c}{$\sigma_{\mr{PN1}}$}\\ 
\hline\noalign{\smallskip}
63.28	&$0.573	_{-0.217 }^{+0.1202} $&$0.1267_{-0.0405}^{+0.0288} $&$1.5873_{-0.1307}^{+0.2493} $&$0.1203_{-0.0386}^{+0.027} $&	& & & \\
\smallskip
71.07	&$0.3826_{-0.0709}^{+0.1115} $&$0.1483_{-0.0059}^{+5.9571} $&$2.8131_{-0.313}^{+0.3377} $&$0.0936_{-0.0088}^{+0.0089} $&	& & & \\
\smallskip
72.23	&$0.2974_{-0.0225}^{+0.0305} $&$0.1356_{-0.0058}^{+0.0057} $&$2.3427_{-0.1301}^{+0.1228} $&$0.1107_{-0.0037}^{+0.0043} $&	& & & \\
\smallskip
73.20	&$0.2648_{-0.0291}^{+0.0238} $&$0.1511_{-0.0073}^{+0.0054} $&$2.3249_{-0.2369}^{+0.1135} $&$0.1122_{-0.0037}^{+5.9924} $&$7.4370_{-0.4220}^{+0.4240} $&	$0.0182_{-0.0023}^{+0.0025}$  &  11.82&	3.96	 \\
\smallskip
74.23	&$0.2717_{-0.0527}^{+0.0863} $&$0.1526_{-0.0166}^{+0.0213} $&$2.5201_{-0.2938}^{+0.2863} $&$0.1066_{-0.0073}^{+0.0103} $&$6.2010_{-0.8120}^{+0.6670} $&	$0.0306_{-0.005}^{+0.0054}$  &   1.16&	3.06	 \\
\smallskip
75.21	&$0.3734_{-0.0979}^{+0.0535} $&$0.1936_{-0.0282}^{+0.0109} $&$2.8001_{-0.3215}^{+0.4232} $&$0.0891_{-0.0092}^{+0.0087} $&$6.3070_{-0.7930}^{+0.6300} $&	$0.0314_{-0.0048}^{+0.0066}$  &  1.17&	3.27	 \\
\smallskip
76.17	&$0.2167_{-0.0197}^{+0.0329} $&$0.1605_{-0.0064}^{+0.0096} $&$2.2066_{-0.2175}^{+0.305} $&$0.1177_{-0.0106}^{+0.0067} $&$5.8600_{-0.6650}^{+0.5410} $&	$0.0344_{-0.0045}^{+0.0068}$  &  1.11&	3.82	 \\
\end{tabular} 
\end{center}
\label{Tab:pds1ni}
Notes:\\ 
rms: root mean square; $\nu_{\mr{max}}$: characteristic frequency;  $\sigma$: significance; BLN: band limited noise; PN: peaked noise
\end{table*}

\addtocounter{table}{-1}
\begin{table*}
\caption{continued}
\begin{center}
\footnotesize
\begin{tabular}{rrrrrrrrr}
\hline\noalign{\smallskip}
 \multicolumn{1}{c}{day} & \multicolumn{1}{c}{$\nu_{\mr{max; BLN1}}$} & \multicolumn{1}{c}{rms$_{\mr{BLN1}}$} & \multicolumn{1}{c}{$\nu_{\mr{max; BLN2}}$} & \multicolumn{1}{c}{rms$_{\mr{BLN2}}$} & \multicolumn{1}{c}{$\nu_{\mr{max; PN1}}$} & \multicolumn{1}{c}{rms$_{\mr{PN1}}$} & \multicolumn{1}{c}{Q$_{\mr{PN1}}$}  & \multicolumn{1}{c}{$\sigma_{\mr{PN1}}$}\\ 
\hline\noalign{\smallskip}
116.65	&$1.6934_{-0.1406}^{+0.2662} $&$0.0544_{-0.0034}^{+0.0031} $&$5.9052_{-0.5089}^{+1.2312} $&$0.0354_{-0.0048}^{+0.0028} $&	& & & \\
\smallskip
116.71	&$1.0966_{-0.1373}^{+0.1637} $&$0.044_{-0.0044}^{+0.0039} $&$6.9146_{-1.077}^{+2.0193} $&$0.0398_{-0.0045}^{+0.0049} $&	& & & \\
\smallskip
116.72	&$1.2204_{-0.2113}^{+0.2894} $&$0.0455_{-0.005}^{+0.0062} $&$5.1461_{-0.8396}^{+1.7196} $&$0.045_{-0.0078}^{+0.0037} $&	& & & \\
\smallskip
116.91	&$1.2657_{-0.2531}^{+0.3836} $&$0.042_{-0.0056}^{+0.0079} $&$5.0449_{-0.7104}^{+1.8626} $&$0.048_{-0.0089}^{+0.0036} $&	& & & \\
\smallskip
116.91	&$-$&$-$&$	2.2086_{-0.297}^{+0.3383} $&$0.0475_{-0.0026}^{+0.0025} $&& & & \\
\smallskip
198.84	&$6.4038_{-0.7449}^{+1.0588} $&$0.0329_{-0.002}^{+0.0008} $&$1.859_{-0.118}^{+0.182} $&$0.0297_{-0.0011}^{+0.0014} $&	& & & \\
\smallskip
199.42	&$0.403	_{-0.044 }^{+0.115} $&$0.0641_{-0.0034}^{+0.0025} $&$3.2843_{-0.2796}^{+0.4732} $&$0.0556_{-0.0036}^{+0.0021} $&	& & & \\
\end{tabular} 
\end{center}
\label{Tab:pds1ni}
Notes:\\ 
rms: root mean square; $\nu_{\mr{max}}$: characteristic frequency;  $\sigma$: significance; BLN: band limited noise; PN: peaked noise
\end{table*}

\addtocounter{table}{-1}
\begin{table*}
\caption{continued}
\begin{center}
\footnotesize
\begin{tabular}{rrrrrrrrr}
\hline\noalign{\smallskip}
 \multicolumn{1}{c}{day} & \multicolumn{1}{c}{$\nu_{\mr{max; BLN1}}$} & \multicolumn{1}{c}{rms$_{\mr{BLN1}}$} & \multicolumn{1}{c}{$\nu_{\mr{max; BLN2}}$} & \multicolumn{1}{c}{rms$_{\mr{BLN2}}$} & \multicolumn{1}{c}{$\nu_{\mr{max; PN1/BLN3}}$} & \multicolumn{1}{c}{rms$_{\mr{PN1/BLN3}}$} & \multicolumn{1}{c}{Q$_{\mr{PN1}}$}  & \multicolumn{1}{c}{$\sigma_{\mr{PN1}}$}\\ 
\hline\noalign{\smallskip}
250.54  &$<0.0148 $&$  0.8296_{-0.3826}^{+0.1704} $&$    0.0678_{-0.0202}^{+0.0272} $&$  0.0542_{-0.0097}^{+0.0112} $&  & & & \\                  
\smallskip
251.45  &$<0.0499 $&$  0.4251_{-0.4251}^{+0.5435} $&$    0.019_{-0.0125}^{+0.0523} $&$  0.0525_{-0.0425}^{+0.036} $&  & & & \\                  
\smallskip
252.34  &$<0.0162 $&$  0.2111_{-0.1704}^{+0.0229} $&  & & & & & \\                            
\smallskip
253.24  &$0.0042_{-0.0042}^{+0.0059} $&$  0.2797_{-0.0948}^{+1.143} $&  & & & & & \\                              
\smallskip
254.20  &$0.1907_{-0.0336}^{+0.0449} $&$  0.1861_{-0.0136}^{+0.0127} $&  & & & & & \\                                
\smallskip
255.83  &$0.0008_{-0.0002}^{+0.0071} $&$  0.9237_{-0.6264}^{+0.0763} $&  & & & & & \\                                
\smallskip
333.97  &$0.0003_{-0.0001}^{+0.0042} $&$  0.9075_{-0.6842}^{+0.0925} $&  & & & & & \\
\end{tabular} 
\end{center}
\label{Tab:pds1ni}
Notes:\\ 
rms: root mean square; $\nu_{\mr{max}}$: characteristic frequency;  $\sigma$: significance; BLN: band limited noise; PN: peaked noise
\end{table*}

\addtocounter{table}{-1}
\begin{table*}
\caption{continued}
\begin{center}
\footnotesize
\begin{tabular}{rrrrrrrrr}
\hline\noalign{\smallskip}
 \multicolumn{1}{c}{day} & \multicolumn{1}{c}{$\nu_{\mr{max; BLN1}}$} & \multicolumn{1}{c}{rms$_{\mr{BLN1}}$} & \multicolumn{1}{c}{$\nu_{\mr{max; BLN2}}$} & \multicolumn{1}{c}{rms$_{\mr{BLN2}}$} & \multicolumn{1}{c}{$\nu_{\mr{max; PN1/BLN3}}$} & \multicolumn{1}{c}{rms$_{\mr{PN1/BLN3}}$} & \multicolumn{1}{c}{Q$_{\mr{PN1}}$}  & \multicolumn{1}{c}{$\sigma_{\mr{PN1}}$}\\ 
\hline\noalign{\smallskip}
529.77	&$0.008_{-0.007}^{+0.018}$ & $	0.214_{-0.063}^{+5.896}$ & $	0.241_{-0.060}^{+0.064}$ & $	0.305_{-0.029}^{+0.024}$ & $	2.031_{-0.748}^{+1.216}$ & $	0.145_{-0.035}^{+0.033}$ &  & \\ 
\smallskip
531.59	&$0.017_{-0.006}^{+0.008}$ & $	0.269_{-0.049}^{+0.038}$ & $	0.291_{-0.106}^{+0.137}$ & $	0.248_{-0.064}^{+0.038}$ & $	1.096_{-0.275}^{+0.445}$ & $	0.225_{-0.055}^{+0.047}$ &  & \\ 
\smallskip
532.75	&$0.011_{-0.004}^{+0.006}$ & $	0.309_{-0.045}^{+0.056}$ & $	0.273_{-0.034}^{+0.045}$ & $	0.329_{-0.015}^{+0.013}$ & $	2.702_{-1.312}^{+2.299}$ & $	0.112_{-0.031}^{+0.035}$ &  & \\ 
\smallskip
533.01	&$0.008_{-0.002}^{+0.002}$ & $	0.325_{-0.026}^{+0.027}$ & $	0.216_{-0.031}^{+0.038}$ & $	0.287_{-0.014}^{+0.012}$ & $	1.732_{-0.292}^{+0.355}$ & $	0.183_{-0.014}^{+0.014}$ &  & \\ 
\smallskip
534.87	&$0.024_{-0.010}^{+0.427}$ & $	0.246_{-0.031}^{+0.029}$ & $	0.282_{-0.063}^{+0.168}$ & $	0.296_{-0.016}^{+0.017}$ & $	2.438_{-0.758}^{+2.562}$ & $	0.159_{-0.036}^{+0.024}$ &  & \\ 
\smallskip
536.68	&$0.008_{-0.002}^{+0.003}$ & $	0.364_{-0.039}^{+0.036}$ & $	0.361_{-0.094}^{+0.098}$ & $	0.263_{-0.018}^{+0.018}$ & $	2.971_{-0.631}^{+0.741}$ & $	0.169_{-0.015}^{+0.017}$ &  & \\ 
\smallskip
537.20	&$0.011_{-0.003}^{+0.004}$ & $	0.291_{-0.024}^{+0.024}$ & $	0.294_{-0.039}^{+0.042}$ & $	0.287_{-0.011}^{+0.010}$ & $	2.319_{-0.425}^{+0.536}$ & $	0.158_{-0.013}^{+0.014}$ &  & \\
\end{tabular} 
\end{center}
\label{Tab:pds1ni}
Notes:\\ 
rms: root mean square; $\nu_{\mr{max}}$: characteristic frequency;  $\sigma$: significance; BLN: band limited noise; PN: peaked noise
\end{table*}

\begin{table*}
\caption{Parameters of the QPOs of the PDS derived from NICER data}
\begin{center}
\begin{tabular}{rrrrrr}
\hline\noalign{\smallskip}
 \multicolumn{1}{c}{day} & \multicolumn{1}{c}{$\nu_{\mr{0; QPO1}}$} & \multicolumn{1}{c}{$\Delta_{\mr{QPO1}}$} & \multicolumn{1}{c}{rms$_{\mr{QPO1}}$} & \multicolumn{1}{c}{Q$_{\mr{QPO1}}$}  & \multicolumn{1}{c}{$\sigma_{\mr{QPO1}}$}\\ 
\hline\noalign{\smallskip}
10.39  &  $0.0344_{-0.0014}^{+0.0019} $  &  $0.0026_{-0.0026}^{+0.0043} $  &  $0.0644_{-0.0139}^{+0.0151} $  &  6.620  &2.320   \\ 
\smallskip
11.99  &  $0.0397_{-0.0009}^{+0.0001} $  &  $0.0000_{-0.0000}^{+0.0026} $  &  $0.0491_{-0.008}^{+0.0065} $  &  7.630  &3.070   \\ 
\smallskip
18.04   &  $0.05_{-0.0026}^{+0.0031} $  &  $0.0090_{-0.0039}^{+0.0054} $  &  $0.0997_{-0.0175}^{+0.0158} $  &  2.780  &2.850   \\ 
\smallskip
19.39   &  $0.058_{-0.0088}^{+0  } $  &  $0.0000_{-0.0000}^{+0.0069} $  &  $0.0777_{-0.0099}^{+0.0091} $  &  4.200  &3.920   \\ 
\smallskip
22.03  &  $0.052_{-0.0028}^{+0.0025} $  &  $0.0151_{-0.0055}^{+0.0063} $  &  $0.1035_{-0.0184}^{+0.0224} $  &  1.720  &2.810   \\ 
\smallskip
23.32  &  $0.0619_{-0.0026}^{+0.0029} $  &  $0.0094_{-0.0037}^{+0.0034} $  &  $0.099_{-0.0138}^{+0.0134} $  &  3.290  &3.590   \\ 
\smallskip
24.02  &  $0.0597_{-0.0032}^{+0.0036} $  &  $0.0172_{-0.0030}^{+0.0052} $  &  $0.099_{-0.0072}^{+0.0055} $  &  1.740  &6.880   \\
\end{tabular} 
\end{center}
\label{Tab:qpo_ni}
Notes:\\ 
rms: root mean square; $\nu_0$: centroid frequency; $\Delta$: half width at half maximum; $\sigma$: significance; QPO: quasiperiodic oscillation
\end{table*}

\addtocounter{table}{-1}
\begin{table*}
\caption{continued}
\begin{center}
\begin{tabular}{rrrrrr}
\hline\noalign{\smallskip}
 \multicolumn{1}{c}{day} & \multicolumn{1}{c}{$\nu_{\mr{0; QPO1}}$} & \multicolumn{1}{c}{$\Delta_{\mr{QPO1}}$} & \multicolumn{1}{c}{rms$_{\mr{QPO1}}$} & \multicolumn{1}{c}{Q$_{\mr{QPO1}}$}  & \multicolumn{1}{c}{$\sigma_{\mr{QPO1}}$}\\ 
\hline\noalign{\smallskip}
71.07   &  $0.4915_{-0.0125}^{+0.021} $  &  $0.0383_{-0.0319}^{+0.0808} $  &  $0.0374_{-0.0099}^{+6.0663} $  &  6.420  &1.890   \\ 
\smallskip
72.23   &  $0.4984_{-0.0088}^{+0.0095} $  &  $0.1042_{-0.0168}^{+0.0194} $  &  $0.0659_{-0.0061}^{+0.0067} $  &  2.390  &5.400   \\ 
\smallskip
73.20   &  $0.4238_{-0.0105}^{+0.0098} $  &  $0.0838_{-0.0175}^{+0.0219} $  &  $0.0602_{-0.007}^{+0.0088} $  &  2.530  &4.300   \\ 
\smallskip
74.23   &  $0.3894_{-0.0212}^{+0.0255} $  &  $0.1004_{-0.0344}^{+0.0552} $  &  $0.0674_{-0.0153}^{+0.0227} $  &  1.940  &2.200   \\ 
\smallskip
75.21   &  $0.3326_{-0.003}^{+0.0296} $  &  $0.0000_{-0.0000}^{+0.0527} $  &  $0.026_{-0.0068}^{+0.0065} $  &  3.160  &1.910   \\ 
\smallskip
76.17   &  $0.3576_{-0.0107}^{+0.0132} $  &  $0.0501_{-0.0203}^{+0.0196} $  &  $0.056_{-0.0118}^{+0.0086} $  &  3.570  &2.370   \\ 
\smallskip
77.06   &  $0.3491_{-0.0094}^{+0.0126} $  &  $0.0619_{-0.0197}^{+0.0331} $  &  $0.0618_{-0.0115}^{+0.0149} $  &  2.820  &2.690   \\
\end{tabular} 
\end{center}
\label{Tab:qpo_ni}
Notes:\\ 
rms: root mean square; $\nu_0$: centroid frequency; $\Delta$: half width at half maximum; $\sigma$: significance; QPO: quasiperiodic oscillation
\end{table*}

\addtocounter{table}{-1}
\begin{table*}
\caption{continued}
\begin{center}
\begin{tabular}{rrrrrr}
\hline\noalign{\smallskip}
 \multicolumn{1}{c}{day} & \multicolumn{1}{c}{$\nu_{\mr{0; QPO1}}$} & \multicolumn{1}{c}{$\Delta_{\mr{QPO1}}$} & \multicolumn{1}{c}{rms$_{\mr{QPO1}}$} & \multicolumn{1}{c}{Q$_{\mr{QPO1}}$}  & \multicolumn{1}{c}{$\sigma_{\mr{QPO1}}$}\\ 
\hline\noalign{\smallskip}
213.35  &  $0.0377_{-0.0055}^{+0.0065} $  &  $0.0213_{-0.0102}^{+0.0164} $  &  $0.1664_{-0.0358}^{+0.0347} $  &  0.880  &2.320   \\ 
\smallskip
214.10  &  $0.0179_{-0.0019}^{+0.0021} $  &  $0.0045_{-0.0022}^{+0.0036} $  &  $0.0954_{-0.0179}^{+0.0172} $  &  1.990  &2.660   \\ 
\smallskip
215.13   &  $0.1739_{-0.0039}^{+0  } $  &  $0.0000_{-0.0000}^{+0.0055} $  &  $0.0484_{-0.0068}^{+0.0059} $  &  15.810  &3.560   \\ 
\smallskip
217.21  &  $0.1495_{-0.0183}^{+0  } $  &  $0.0000_{-0.0000}^{+0.0094} $  &  $0.0556_{-0.0129}^{+0.0104} $  &  7.950  &2.160   \\ 
\smallskip
218.43  &  $0.3326_{-0.0062}^{+0.0029} $  &  $0.0000_{-0.0000}^{+0.0164} $  &  $0.0306_{-0.0074}^{+0.006} $  &  10.140  &2.070   \\ 
\smallskip
222.85  &  $0.1122_{-0.0041}^{+0.0052} $  &  $0.0077_{-0.0077}^{+0.0069} $  &  $0.0763_{-0.0182}^{+0.0162} $  &  7.290  &2.100   \\ 
\smallskip
225.04  &  $0.0458_{-0.0001}^{+0.0122} $  &  $0.0000_{-0.0000}^{+0.0029} $  &  $0.064_{-0.0173}^{+0.0143} $  &  7.970  &1.850   \\
\end{tabular} 
\end{center}
\label{Tab:qpo_ni}
Notes:\\ 
rms: root mean square; $\nu_0$: centroid frequency; $\Delta$: half width at half maximum; $\sigma$: significance; QPO: quasiperiodic oscillation
\end{table*}

\addtocounter{table}{-1}
\begin{table*}
\caption{continued}
\begin{center}
\begin{tabular}{rrrrrr}
\hline\noalign{\smallskip}
 \multicolumn{1}{c}{day} & \multicolumn{1}{c}{$\nu_{\mr{0; QPO2}}$} & \multicolumn{1}{c}{$\Delta_{\mr{QPO2}}$} & \multicolumn{1}{c}{rms$_{\mr{QPO2}}$} & \multicolumn{1}{c}{Q$_{\mr{QPO2}}$}  & \multicolumn{1}{c}{$\sigma_{\mr{QPO2}}$}\\ 
\hline\noalign{\smallskip}
10.39   &  $0.0779_{-0.0084}^{+0.0103}$  &$  0.0261_{-0.0060}^{+0.0087}$  &$  0.0902_{-0.0116}^{+0.0096}$  &  1.49  &3.89   \\ 
\smallskip
18.04   &  $0.1315_{-0.017}^{+0.0119}$  &$  0.0509_{-0.0141}^{+0.0179}$  &$  0.1353_{-0.0234}^{+0.0375}$  &  1.29  &2.89 \\ 
\smallskip
19.39   &  $0.1197_{-0.0071}^{+0.0078}$  &$  0.0601_{-0.0114}^{+0.0081}$  &$  0.1731_{-0.0267}^{+0.0144}$  &  1.00  &3.24 \\ 
\smallskip
22.03   &  $0.1219_{-0.0092}^{+0.0073}$  &$  0.0876_{-0.0060}^{+0.0047}$  &$  0.1778_{-0.0139}^{+0.011}$   &  0.70  &6.40 \\ 
\smallskip
23.32   &  $0.1343_{-0.0039}^{+0.0064}$  &$    0.011_{-0.011}^{+0.0106}$  &$  0.0612_{-0.0158}^{+0.0216}$   &  6.10  &1.94 \\ 
\smallskip
24.02   &  $0.149_{-0.0016}^{+0.0015}$  &$  0.0027_{-0.0027}^{+0.0075}$  &$  0.0454_{-0.0088}^{+0.0108}$  &  27.59  &2.58   \\ 
\smallskip
25.31   &  $0.1756_{-0.0095}^{+0.0114}$  &$  0.0451_{-0.0063}^{+0.0138}$  &$  0.0987_{-0.0059}^{+0.0104}$  &  1.95  &8.36   \\
\end{tabular} 
\end{center}
\label{Tab:qpo_ni}
Notes:\\ 
rms: root mean square; $\nu_0$: centroid frequency; $\Delta$: half width at half maximum; $\sigma$: significance; QPO: quasiperiodic oscillation
\end{table*}

\addtocounter{table}{-1}
\begin{table*}
\caption{continued}
\begin{center}
\begin{tabular}{rrrrrr}
\hline\noalign{\smallskip}
 \multicolumn{1}{c}{day} & \multicolumn{1}{c}{$\nu_{\mr{0; QPO2}}$} & \multicolumn{1}{c}{$\Delta_{\mr{QPO2}}$} & \multicolumn{1}{c}{rms$_{\mr{QPO2}}$} & \multicolumn{1}{c}{Q$_{\mr{QPO2}}$}  & \multicolumn{1}{c}{$\sigma_{\mr{QPO2}}$}\\ 
\hline\noalign{\smallskip}
75.21   &  $0.7351_{-0.0292}^{+0.0002}$  &$  0.0005_{-0.0005}^{+0.0423}$  &$  0.025_{-0.0064}^{+0.0049}$  &  8.69  &1.95   \\ 
\smallskip
76.17   &  $0.6835_{-0.0259}^{+0.0178}$  &$  0.0454_{-0.0454}^{+0.0339}$  &$  0.0437_{-0.0126}^{+0.0132}$  &  7.53  &1.73   \\ 
\smallskip
77.06   &  $0.694_{-0.0379}^{+0.0167}$  &$  0.0509_{-0.0382}^{+0.0437}$  &$  0.0405_{-0.0104}^{+6.0632}$  &  6.82  &1.95   \\ 
\smallskip
78.22   &  $0.7332_{-0.0173}^{+0.0119}$  &$  0.0636_{-0.042}^{+0.037}$  &$    0.0578_{-0.0131}^{+0.0123}$  &  5.76  &2.21   \\ 
\smallskip
79.25   &  $0.8647_{-0.0203}^{+0.0203}$  &$  0.077_{-0.043}^{+0.0434}$  &$  0.0423_{-0.0153}^{+0.0139}$  &  5.61  &1.38   \\ 
\smallskip
80.21   &  $0.859_{-0.0259}^{+0.027}$  &$    0.1884_{-0.0554}^{+0.0674}$  &$  0.0745_{-0.0123}^{+0.012}$   &  2.28  &3.03   \\ 
\smallskip
81.24   &  $0.7353_{-0.0054}^{+0.0372}$  &$  0.0021_{-0.0021}^{+0.0653}$  &$  0.0329_{-0.0045}^{+6.0708}$  &  5.63  &3.66   \\
\end{tabular} 
\end{center}
\label{Tab:pds2nic}
Notes:\\ 
rms: root mean square; $\nu_0$: centroid frequency; $\Delta$: half width at half maximum; $\sigma$: significance; QPO: quasiperiodic oscillation
\end{table*}

\addtocounter{table}{-1}
\begin{table*}
\caption{continued}
\begin{center}
\begin{tabular}{rrrrrr}
\hline\noalign{\smallskip}
 \multicolumn{1}{c}{day} & \multicolumn{1}{c}{$\nu_{\mr{0; QPO2}}$} & \multicolumn{1}{c}{$\Delta_{\mr{QPO2}}$} & \multicolumn{1}{c}{rms$_{\mr{QPO2}}$} & \multicolumn{1}{c}{Q$_{\mr{QPO2}}$}  & \multicolumn{1}{c}{$\sigma_{\mr{QPO2}}$}\\ 
\hline\noalign{\smallskip}
199.42  &  $0.1984_{-0  }^{+0.0029}$  &$  0.0000_{-0.0000}^{+0.0056}$  &$  0.0297_{-0.0038}^{+0.0034}$  &  17.71  &3.91   \\ 
\smallskip
202.00  &  $4.4019_{-0.1738}^{+0.1355}$  &$  2.1669_{-0.1278}^{+0.1885}$  &$  0.0558_{-0.0027}^{+0.0042}$  &  1.02  &10.33   \\ 
\smallskip
203.22  &  $0.4144_{-0.0139}^{+0.0119}$  &$  0.0171_{-0.0171}^{+0.0323}$  &$  0.0275_{-0.0072}^{+0.0064}$  &  12.12  &1.91   \\ 
\smallskip
222.85  &  $0.2352_{-0.008}^{+0.0082}$  &$  0.0148_{-0.0139}^{+0.0191}$  &$  0.0582_{-0.0128}^{+0.0112}$  &  7.95  &2.27   \\ 
\smallskip
231.73  &  $0.0336_{-0.006}^{+0  }$  &$    0.0000_{-0.0000}^{+0.0009}$  &$  0.0795_{-0.0239}^{+0.0182}$  &  18.85  &1.66   \\ 
\smallskip
232.06  &  $0.04_{-0.02  }^{+0  }$  &$    0.3058_{-0.0854}^{+0.1038}$  &$  0.126_{-0.0191}^{+0.0157}$  &  0.07  &3.30   \\ 
\smallskip
233.09  &  $0.0336_{-0  }^{+0.0053}$  &$  0.0000_{-0.0000}^{+0.0012}$  &$  0.0369_{-0.01  }^{+0.0085}$  &  13.53  &1.85   \\
\end{tabular} 
\end{center}
\label{Tab:pds2nic}
Notes:\\ 
rms: root mean square; $\nu_0$: centroid frequency; $\Delta$: half width at half maximum; $\sigma$: significance; QPO: quasiperiodic oscillation
\end{table*}

\begin{table*}
\caption{Additional Lorentzian needed to fit QPO at higher frequency in NICER data}
\begin{center}
\begin{tabular}{rrrrrr}
\hline\noalign{\smallskip}
 \multicolumn{1}{c}{day} & \multicolumn{1}{c}{$\nu_{\mr{0}}$} & \multicolumn{1}{c}{$\Delta$} & \multicolumn{1}{c}{rms} & \multicolumn{1}{c}{Q}  & \multicolumn{1}{c}{$\sigma$}\\ 
\hline\noalign{\smallskip}
23.32  &  $0.1601_{-0.0203}^{+0.0278}$  &$0.0719_{-0.0199}^{+0.0360}$  &$0.1221\pm0.0357$  &  1.11  &1.71  \\ 
\smallskip
24.02  &  $0.1938_{-0.0095}^{+0.0156}$  &$0.0614_{-0.0070}^{+0.0087}$  &$0.094_{-0.0082}^{+0.0063}$  &  1.58  &5.73  \\ 
\smallskip
36.07  &  $0.3077_{-0.0081}^{+0.0078}$  &$0.0480_{-0.0101}^{+0.0109}$  &$0.0592_{-0.0045}^{+0.0038}$  &  3.21  &6.58  \\ 
\smallskip
52.23  &  $0.6216_{-0.0584}^{+0.0725}$  &$0.3772_{-0.0327}^{+0.0502}$  &$0.1172_{-0.0138}^{+0.0126}$  &  0.82  &4.25  \\ 
\smallskip
53.24  &  $0.6408_{-0.0634}^{+0.1074}$  &$0.3823_{-0.0500}^{+0.0422}$  &$0.1137_{-0.0228}^{+0.0131}$  &  0.84  &2.49\\ 
\smallskip
54.27  &  $0.8235_{-0.0483}^{+0.0529}$  &$0.2084_{-0.0778}^{+0.0899}$  &$0.0533_{-0.0142}^{+0.0206}$  &  1.98  &1.88\\ 
\smallskip
55.24  &  $0.8462_{-0.0719}^{+0.1025}$  &$0.4123_{-0.0578}^{+0.0578}$  &$0.0914_{-0.0157}^{+0.0128}$  &  1.03  &2.91\\
\end{tabular} 
\end{center}
\label{Tab:pds2nic}
Notes:\\ 
rms: root mean square; $\nu_0$: centroid frequency; $\Delta$: half width at half maximum; $\sigma$: significance; QPO: quasiperiodic oscillation
\end{table*}

\begin{table*}
\caption{Additional QPOs or peaked noise features}
\begin{center}
\begin{tabular}{rrrrrr}
\hline\noalign{\smallskip} 
 \multicolumn{1}{c}{day} & \multicolumn{1}{c}{$\nu_{\mr{0}}$} & \multicolumn{1}{c}{$\Delta$} & \multicolumn{1}{c}{rms}   &  \multicolumn{1}{c}{Q}  & \multicolumn{1}{c}{$\sigma$}  \\ 
\hline\noalign{\smallskip}
85.26   &  $0.1312_{-0.0056}^{+0.0011}$  &$<0.0138$  &$0.032_{-0.0066}^{+0.0103}$  &  4.75  &2.42  \\ 
\smallskip
93.17   &  $1.4008_{-0.0142}^{+0.0076}$  &$<0.0556$  &$0.0241_{-0.0043}^{+0.0036}$  &  12.59  &2.80  \\ 
\smallskip
96.24   &  $0.5_{-0.0225}^{+0.0183}$  &$0.3001_{-0.0126}^{+0.0152}$  &$0.1524_{-0.003}^{+0.0031}$  &  0.83  &25.40  \\ 
\smallskip
116.65  &  $0.5009_{-0.0306}^{+0.0252}$  &$0.372_{-0.054}^{+0.0814}$  &$0.0328_{-0.0034}^{+0.0043}$  &  3.08  &4.82  \\ 
\smallskip
116.71  &  $0.6799_{-0.0604}^{+0.043}$  &$0.3278_{-0.0679}^{+0.0858}$  &$0.0263_{-0.0035}^{+0.004}$  &  3.96  &3.76  \\ 
\smallskip
525.05	&  $0.087_{-0.006}^{+0.006}$  &$0.063_{-0.014}^{+0.007}$  &$0.215_{-0.020}^{+0.007}$  &	0.69&	5.29		\\
\hline\noalign{\smallskip}
\end{tabular} 
\end{center}
\label{Tab:pds2nic}
Notes:\\ 
rms: root mean square; $\nu_0$: centroid frequency; $\Delta$: half width at half maximum; $\sigma$: significance; QPO: quasiperiodic oscillation
\end{table*}



\end{document}